\documentclass[fleqn,usenatbib,useAMS]{mnras}

\usepackage{graphicx}	% Including figure files
\usepackage{amsmath}	% Advanced maths commands
\usepackage{amssymb}	% Extra maths symbols
\usepackage{multicol}        % Multi-column entries in tables
\usepackage{bm}		% Bold maths symbols, including upright Greek
\usepackage{pdflscape}	% Landscape pages

\newcommand{\fref}{Fig. \ref}
\newcommand{\tref}{Table \ref}
\usepackage[T1]{fontenc}
\usepackage{ae,aecompl}

\usepackage{newtxtext,newtxmath}
\title[Imaging sub-Jupiter mass exoplanets with JWST]{Direct imaging of sub-Jupiter mass exoplanets with \textit{James Webb Space Telescope} coronagraphy}

\author[A. L. Carter et al.]{Aarynn L. Carter$^{1,2}$\thanks{E-mail: \href{mailto:aarynn.carter@ucsc.edu}{aarynn.carter@ucsc.edu}}, 
Sasha Hinkley$^{1}$,
Mariangela Bonavita$^{3,4}$,
Mark W. Phillips$^{1}$,
\newauthor
Julien H. Girard$^{5}$,
Marshall Perrin$^{5}$,
Laurent Pueyo$^{5}$,
Arthur Vigan$^{6}$,
Jonathan Gagn\'e$^{7}$,
\newauthor
Andrew J. I. Skemer$^{2}$
\\
\\
$^{1}$Astrophysics Group, University of Exeter, Physics Building, Stocker Road, Devon, EX4 4QL, UK \\
$^{2}$University of California, Santa Cruz, Santa Cruz, CA 95064, USA \\
$^{3}$SUPA, Institute for Astronomy, University of Edinburgh, Blackford Hill, Edinburgh EH9 3HJ, UK \\
$^{4}$Centre for Exoplanet Science, University of Edinburgh, Edinburgh EH9 3HJ, UK \\
$^{5}$Space Telescope Science Institute, 3700 San Martin Dr, Baltimore MD, 21218, USA \\
$^{6}$Aix Marseille Univ, CNRS, CNES, LAM, Marseille, France \\
$^{7}$Plan\'etarium Rio Tinto Alcan, Espace pour la Vie, 4801 av. Pierre-de Coubertin, Montr\'eal, Qu\'ebec, Canada}
\date{Last updated XXX; in original form YYY}

\pubyear{2020}

\begin{document}
\label{firstpage}
\pagerange{\pageref{firstpage}--\pageref{lastpage}}
\maketitle

\begin{abstract}
The \textit{James Webb Space Telescope} (\textit{JWST}), currently scheduled to launch in 2021, will dramatically advance our understanding of exoplanetary systems with its ability to \textit{directly} image and characterise planetary-mass companions at wide separations through coronagraphy. Using state-of-the-art simulations of \textit{JWST} performance, in combination with the latest evolutionary models, we present the most sophisticated simulated mass sensitivity limits of \textit{JWST} coronagraphy to date. In particular, we focus our efforts towards observations of members within the nearby young moving groups $\beta$~Pictoris and TW~Hya. These limits indicate that whilst \textit{JWST} will provide little improvement towards imaging exoplanets at short separations, at wide separations the increase in sensitivity is dramatic. We predict \textit{JWST} will be capable of imaging sub-Jupiter mass objects beyond $\sim$30~au, sub-Saturn mass objects beyond $\sim$50~au, and that beyond $\sim$100 au, \textit{JWST} will be capable of directly imaging companions as small as 0.1~$M_\textrm{J}$ $-$ at least an order of magnitude improvement over the leading ground-based instruments. Probing this unexplored parameter space will be of immediate value to modelling efforts focused on planetary formation and population synthesis. \textit{JWST} will also serve as an excellent complement to ground based observatories through its unique ability to characterise previously detected companions across the near- to mid-infrared for the first time. 
\end{abstract}

\begin{keywords}
infrared: planetary systems -- planets and satellites: gaseous planets -- planets and satellites: atmospheres -- techniques: high angular resolution -- techniques: image processing
\end{keywords}

%%%%%%%%%%%%%%%%%%%%%%%%%%%%%%%%%%%%%%%%%%%%%%%%%%

%%%%%%%%%%%%%%%%% BODY OF PAPER %%%%%%%%%%%%%%%%%%
\section{Introduction}\label{sec:intro}
The direct imaging of exoplanetary companions remains a critical avenue towards our understanding of planetary formation and evolution due to its ability to probe the widest separations of exoplanetary systems, a region of parameter space largely inaccessible to the more prolific transit or radial velocity methods. Over the last decade, direct imaging surveys have optimised their target selection to maximise the detection of exoplanetary-mass companions (e.g. \citealt{Niel13, Viga17, Ston18, Niel19}). In particular, focus is often placed on the distance and age of the target systems. When a star is closer to us, the physical separations explored in an observation correspond to larger angular separations, meaning the emission of a companion object will be further apart from the considerably brighter stellar emission and therefore easier to detect. Additionally, it is at the shortest physical separations (particularly $\lesssim$50~au) that planetary-mass companions more commonly occur \citep{Mord18, Emse20}. Younger stars are preferable targets as at early ages exoplanets are more luminous owing to their more recent formation, and are therefore easier to detect \citep{Burr97, Bara03, Phil20}. Furthermore, at the earliest ages it is much easier to differentiate between potential scenarios for the initial entropy conditions of an exoplanet and in turn more robustly measure its mass \citep{Marl07, Spie12, Marl14}. Whilst identifying the nearest stellar systems is relatively straightforward, obtaining precise estimates of their ages is particularly challenging \citep{Mama08, Sode10}, and significantly limits our ability to select optimal survey samples. 

Fortunately, this limitation can be overcome by selectively observing objects within nearby young moving groups, coeval associations of stars that share the same galactic space motion. These associations provide a unique advantage, as the ages of their constituent members can be determined more robustly (within a few Myr) by combining typical indicators of youth with their 3-dimensional galactic motions \citep{Bell15}. Young moving groups remain a principal area of exploration towards directly imaging exoplanets, and a number of surveys have focused their attention at least partially towards them (e.g. \citealt{Lafr07, Bill13, Bran14, Chau15, Bowl15, Gali16, Niel19}). In fact, a large proportion of the known directly imaged exoplanets have been discovered around stars within young moving groups \citep{Bowl16}.

Discovering and further characterising exoplanetary companions through direct imaging is most readily accomplished with near- to mid-infrared observations. It is at these wavelengths that the spectral energy distributions of planetary-mass objects peak and the contrast between companion and host star is at its lowest \citep{Skem14}. Additionally, infrared wavelengths are rich with spectral absorption features that enable us to probe atmospheric structure, dynamics, and composition, in addition to the overall formation and migration history of an exoplanet \citep{Madh19}. The overwhelming majority of direct imaging surveys have been performed from the ground~$-$~where it is feasible to build a telescope with a spatial resolving power large enough at these wavelengths to detect exoplanetary companions. However, these observations are not without limitations. The effect of Earth's atmosphere is significant: advanced adaptive optics techniques must be used to account for atmospheric wavefront distortion, only particular wavelength regions can be observed due to the atmospheres inherent transmittance, and day-to-day weather variations reduce observing efficiency. Furthermore, the increasing noise resulting from the thermal emissivity of the telescope itself compounds with the intrinsically larger sky thermal background beyond $\sim$3 $\mu$m. It has therefore become more and more desirable to conduct these observations from space.

Currently scheduled for launch during 2021, the \textit{James Webb Space Telescope} (\textit{JWST}) \citep{Gard06} will significantly transform our ability to both detect and characterise exoplanets through direct imaging. \textit{JWST} will be host to a diverse range of observing modes across its four instruments, enabling a similarly diverse range of observations to be performed. Crucially, \textit{JWST} will have the largest aperture of any space telescope to date, bypassing all of the ground-based concerns of the Earth's atmosphere and allowing it to reach an unprecedented level of sensitivity. Furthermore, as \textit{JWST} is located in space it is able to have a very broad functional wavelength range, spanning from $\sim$0.6$-$28~$\mu$m. When considering only the observing modes relevant to direct imaging of the closest separation exoplanets, this range is reduced to $\sim$1$-$16~$\mu$m. However, this is still a significant increase over current instruments which are constrained below $\sim$5~$\mu$m.

Given the relatively short amount of time until the launch of \textit{JWST}, it is prudent to assess the predicted capabilities of its direct imaging modes towards exoplanet detection. Such an analysis was initially performed by \citet{Beic10}, however the understanding of \textit{JWST} performance has increased significantly over the last decade due to a variety of observatory tests and the creation of robust simulation tools. Furthermore, new and sophisticated exoplanet atmosphere and evolutionary models have been produced (e.g. \citealt{Lind19, Phil20}), enabling a more accurate determination of their predicted luminosities. Recently, \citet{Perr18} provided a significant, target unspecific, update to the contrast predictions for \textit{JWST}, \cite{Dani18} calculated the detectability of a sub-sample of the known directly imaged exoplanet population with \textit{JWST} mid-infrared coronagraphy and spectroscopy, \citet{Sall19} gave an account of the non-redundant masking and kernel phase capabilities of \textit{JWST}, and finally \citet{Bran20} have explored the feasibility of \textit{JWST} mid-infrared coronagraphic imaging of field-aged exoplanets around the nearest stars. With these studies in mind, we instead focus our efforts towards determining the overall detection limits of \textit{JWST} coronagraphic imaging towards sub-Jupiter mass exoplanets for a set of observations of the previously discussed nearby young moving groups. Sensitivity to such low mass objects is largely unattainable with current instrumentation, and will be a unique advantage of \textit{JWST} coronagraphy as demonstrated in this study. 

In Section \ref{sec:ymg_selection} we outline our choice of young moving groups and the objects within them, in Section \ref{sec:simulations} we describe the performed simulations, and in Section \ref{sec:results} we describe the conversion of these simulations to detection probability maps. Our primary results are discussed in Section \ref{sec:discussion}, and finally we summarise our conclusions in Section \ref{sec:conclusions}
%%%%%%%%%%%%%%%%%%%%%%%%%%%%%%%%%%%%%%%%%%%%%%%%%%%%%%%%%%%%%%%%%%%%%%%%%%%%%%%%%%%%%%%%%%%%%%%%%%%%%%%%%
%%% YMG SELECTION
%%%%%%%%%%%%%%%%%%%%%%%%%%%%%%%%%%%%%%%%%%%%%%%%%%%%%%%%%%%%%%%%%%%%%%%%%%%%%%%%%%%%%%%%%%%%%%%%%%%%%%%%%
\section{Young Moving Group Selection}\label{sec:ymg_selection}
As there are currently 26 known, well defined, associations younger than $\sim$ 200 Myr within 150~pc \citep{Gagn18a, Gagn18c, Zuck19, Mein19, Curt19}, it is necessary to select the young moving groups among this sample that will be best suited for searches of wide separation companions through direct imaging. Specifically, throughout this study we focus our efforts on the TW~Hya Association (TWA) \citep{Kast97, Gagn18a} and the $\beta$~Pictoris Moving Group ($\beta$PMG) \citep{Zuck01, Gagn18a}. Both of these moving groups occupy a unique region of parameter space, with ages old enough that planetary formation processes have largely ended due to disk clearing \citep{Hais01}, ages young enough that any potentially formed planets have retained a significant amount of heat from their initial gravitational contraction and are therefore more luminous \citep{Bara03, Phil20}, and distances close enough to more favourably probe the innermost architectures of planetary systems through direct imaging. Although many other moving groups fulfil one or even two of these qualities \citep{Gagn18a}, currently TWA and $\beta$PMG present the best opportunity to fulfil all three at once and are hence chosen for this investigation. 

Whilst there is a known distribution in the distances of individual young moving group members, generally the members of TWA lie $\sim$60~pc away, whereas the members of $\beta$PMG lie $\sim$30~pc away. Therefore, observations of $\beta$PMG members are more likely to probe smaller physical separations where planetary formation is more common \citep{Mord18}. However, TWA has an estimated age of $10 \pm 3$~Myr, whereas $\beta$PMG has an estimated age of $24 \pm 3$~Myr \citep{Malo14, Bell15}. Observations of TWA members are therefore likely to be sensitive to lower mass exoplanets, as they will have formed more recently and be naturally more luminous \citep{Bara03, Phil20}. 

\begin{figure}
    \centering
    \includegraphics[width=\columnwidth]{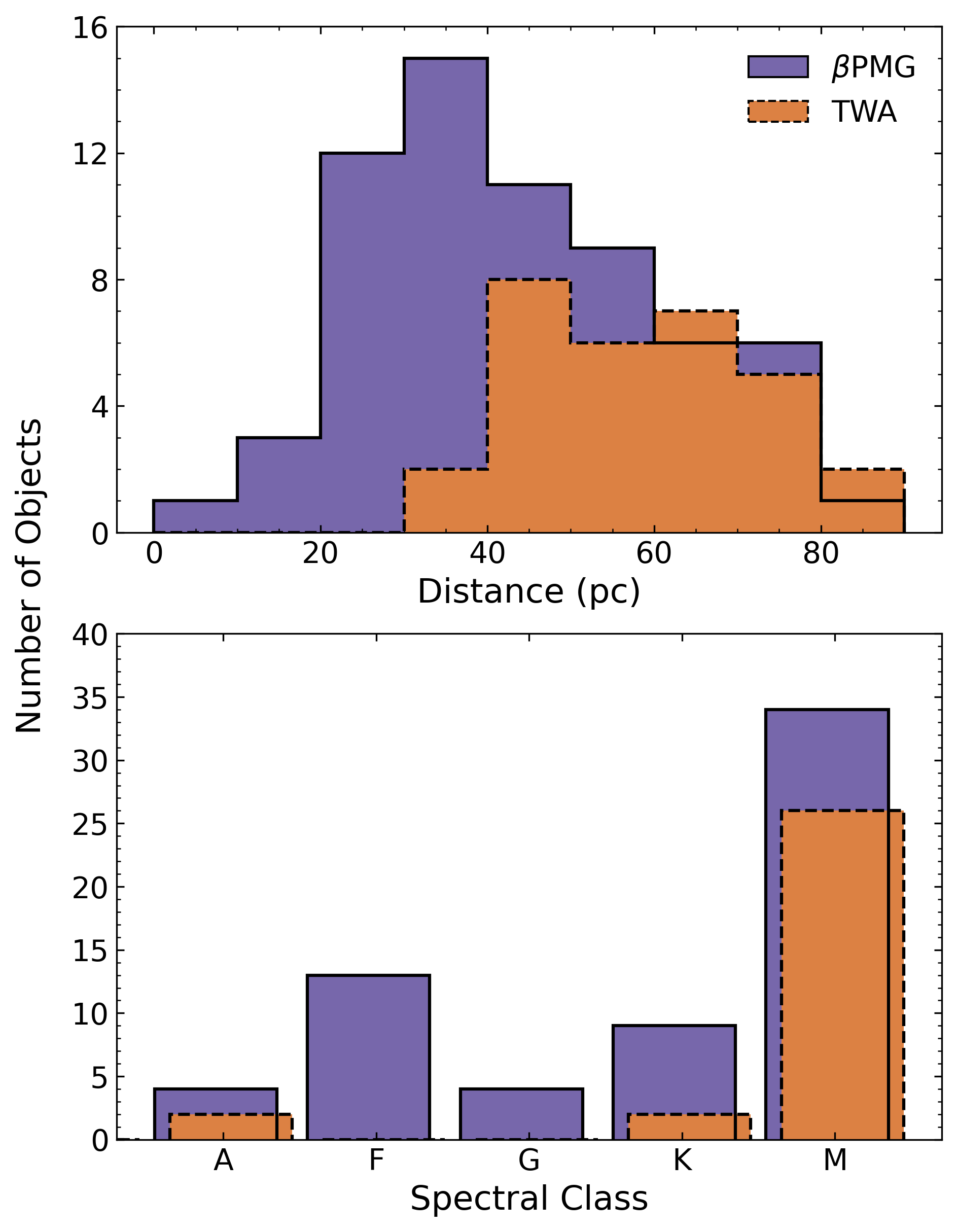}
    \caption{Overlapping histograms of the selected $\beta$PMG (purple, solid line) and TWA (orange, dashed line) populations in both distance (\textit{top}) and in spectral class (\textit{bottom}).}
    \label{fig:sample_hists}
\end{figure}

As the known populations of both TWA and $\beta$PMG have grown since their original classifications, we use the on-going compilation of young association members from the BANYAN tool described in \citet{Gagn18a} (updated with \textit{Gaia} DR2 astrometry and kinematics from \citealt{Gaia16, Gaia18}) to select objects for this study. Specifically, we select objects that: have a high BANYAN membership probability of >90\%, have at least two complimentary measurements from radial velocity, parallax, or youth, and are the primary object within their respective system. Following this selection, we obtain 30~sample objects for TWA and 64~sample objects for $\beta$PMG. Each individual object and its properties are listed in Tables \ref{tab:bpmg_sample} and \ref{tab:twa_sample}, and the distributions of these samples in distance and spectral type are shown in \fref{fig:sample_hists}.

%%%%%%%%%%%%%%%%%%%%%%%%%%%%%%%%%%%%%%%%%%%%%%%%%%%%%%%%%%%%%%%%%%%%%%%%%%%%%%%%%%%%%%%%%%%%%%%%%%%%%%%%%
%%% SIMULATIONS
%%%%%%%%%%%%%%%%%%%%%%%%%%%%%%%%%%%%%%%%%%%%%%%%%%%%%%%%%%%%%%%%%%%%%%%%%%%%%%%%%%%%%%%%%%%%%%%%%%%%%%%%%
\section{\textit{JWST} Coronagraphic Imaging Simulations}\label{sec:simulations}
\begin{figure*}
    \centering
    \includegraphics[width=\textwidth]{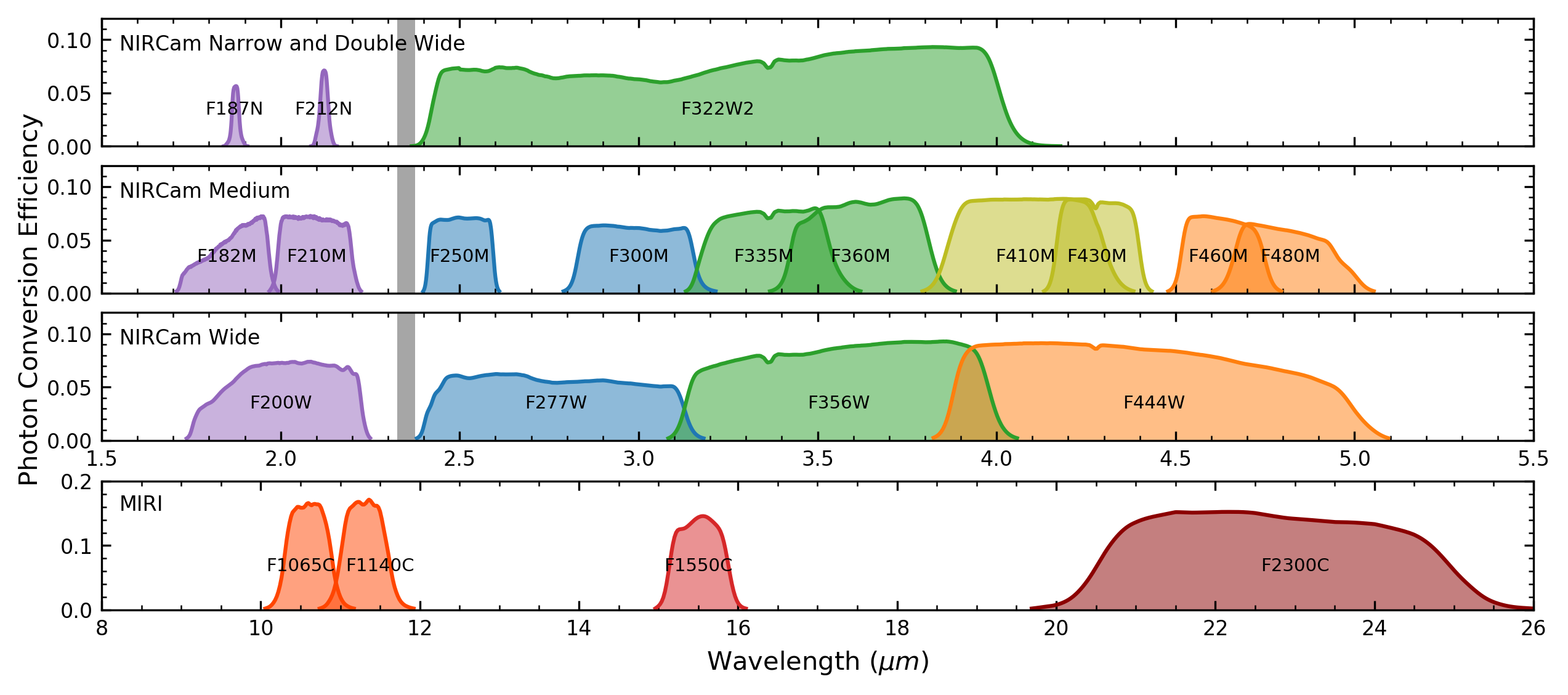}
    \caption[JWST coronagraphic photon conversion efficiencies]{All \textit{JWST} filters that are compatible with its coronagraphic imaging modes and the corresponding photon conversion efficiencies (PCEs) for an observation in such a setup. Grey lines indicate the gap in wavelength coverage between the short and long wavelength channels of the NIRCam detectors. All short wavelength channel NIRCam PCEs are computed using the throughput of the MASK210R round mask, and all long wavelength channel NIRCam PCEs are computed using the MASK335R round mask. All PCEs are determined using the \textit{JWST} exposure time calculator, \texttt{Pandeia} \citep{Pont16}.}
    \label{fig:all_pces}
\end{figure*}

Of the four instruments aboard \textit{JWST}, three have high-contrast imaging capabilities. Namely, the Near-InfraRed Camera (NIRCam) \citep{Riek05} has five separate coronagraphic masks, the Mid-InfraRed Imager (MIRI) \citep{Riek15} has four coronagraphic masks, and the Near-InfraRed Imager and Slitless Spectrograph (NIRISS) \citep{Doyo12} has aperture masking interferometry capabilities through the use of a non-redundant mask \citep{Siva12}. For NIRCam and MIRI in particular, we show the photon conversion efficiencies (PCEs) for all of the filters that can be paired with these masks as calculated by the \textit{JWST} exposure time calculator \texttt{Pandeia} \citep{Pont16} in \fref{fig:all_pces}. These PCEs describe the fraction of incoming photons that will be detected in the final science field of view, and includes the effects of the optical telescope element, detector efficiency, filter throughput, and coronagraph transmission. 

Of the five NIRCam masks, two are of a tapered bar design and span the entire field of view when used, whilst three are of a round design and only obscure a central region of the field of view. Due to their tapered design the bar masks offer superior contrast limits to the round masks at the shortest separations, as a star can be placed at a specific position behind the mask such that the width of the mask at that position best obscures the extent of the stellar PSF in the observed filter of choice. Such capability is ideal for characterisation studies of known objects, however, when the position of a companion object is unknown it is possible that it may lie underneath a bar mask due to its broad span. To avoid such uncertainties, we do not consider the performance of the bar masks in this work. Additionally, we do not consider the MASK210R round mask as it is only compatible with filters below 2.3~$\mu$m, a wavelength range that ground-based telescopes are likely to have superior performance across. Of the two remaining round masks we select the MASK335R for all of the NIRCam simulations due to its smaller inner working angle (IWA) compared to the MASK430R and with the knowledge that any potential improvement in contrast with the MASK430R is relatively small following PSF subtraction \citep{Perr18}. As there are a large number of filters compatible with NIRCam coronagraphic imaging (see \fref{fig:all_pces}), we only select the broad F356W and F444W filters to be simulated in this study. The primary trade-off when selecting a filter is that broader filters have a higher throughput and therefore an improved observational efficiency, but narrower filters can be more focused towards the region of wavelength space where the contrast between the star and a companion is lowest. From a subset of preliminary simulations similar to those described in Section \ref{sec:pcake}, we find that the F444W filter is generally the best suited of the NIRCam filters to identify the lowest mass objects. In a small number of cases however, the F460M and F480M filters perform just as well, if not slightly better, with a $\sim$0.02~$M_\textrm{J}$ greater mass sensitivity. The F444W filter is selected in particular as we are primarily interested in the broad, population based, sensitivity limits of \textit{JWST} coronagraphic imaging, and not those of a small subset of targets. The F356W filter lies directly on a broad CH$_4$ absorption band and is therefore much less suitable for direct imaging of the lowest mass companions. Nevertheless, we find that it is the most optimal filter between $3-4$~$\mu$m and select to include it as a comparison to the F444W filter. 

Three of the four MIRI coronagraphic masks are of a four-quadrant phase mask (4QPM) design at 10.65, 11.40 and 15.50~$\mu$m, with the final mask of a classical Lyot design at 23~$\mu$m \fref{fig:all_pces}. The mask at 23~$\mu$m has a large IWA of 2.16$\arcsec$ which makes it typically unsuitable for exoplanet observations and is therefore not considered as part of this work. All three of the 4QPMs utilise a specific paired filter to optimise the cancellation of stellar light at the centre of the mask and the choice of mask is therefore tied to the wavelength of interest. For this investigation, we select the F1140C and F1550C mask/filter combinations. Only the F1140C filter is selected of the two filters between $10-12$~$\mu$m as the F1065C filter is designed to probe an NH$_3$ absorption band. Such absorption will reduce the received flux and therefore limit the ability to detect cooler (lower mass) exoplanets as NH$_3$ is more abundant in their atmospheres due to the more favourable conversion of N$_2$ to NH$_3$ towards lower temperatures.

The method of direct imaging performed by NIRISS is vastly different to that performed by traditional coronagraphy. NIRISS utilises a non-redundant mask to convert the full \textit{JWST} aperture into an interferometric array. This allows NIRISS to perform observations at a higher angular resolution than \textit{JWST} coronagraphy, but only at separations shorter than $\sim$400~mas \citep{Siva09}. Given the significantly different method of operation versus that of standard coronagraphy and the unique techniques required for their simulation, we do not incorporate any NIRISS observations into this study. For a recent account of NIRISS detection limits, we refer the reader to \citet{Sall19}. 

%%%%%%%%%%%%%%%%%%%%%%%%%%%%%%%%%%%%%%%%%%%%%%%%%%%%
\subsection{SED Selection}\label{sec:sed_select}
\begin{figure}
    \centering
    \includegraphics[width=\columnwidth]{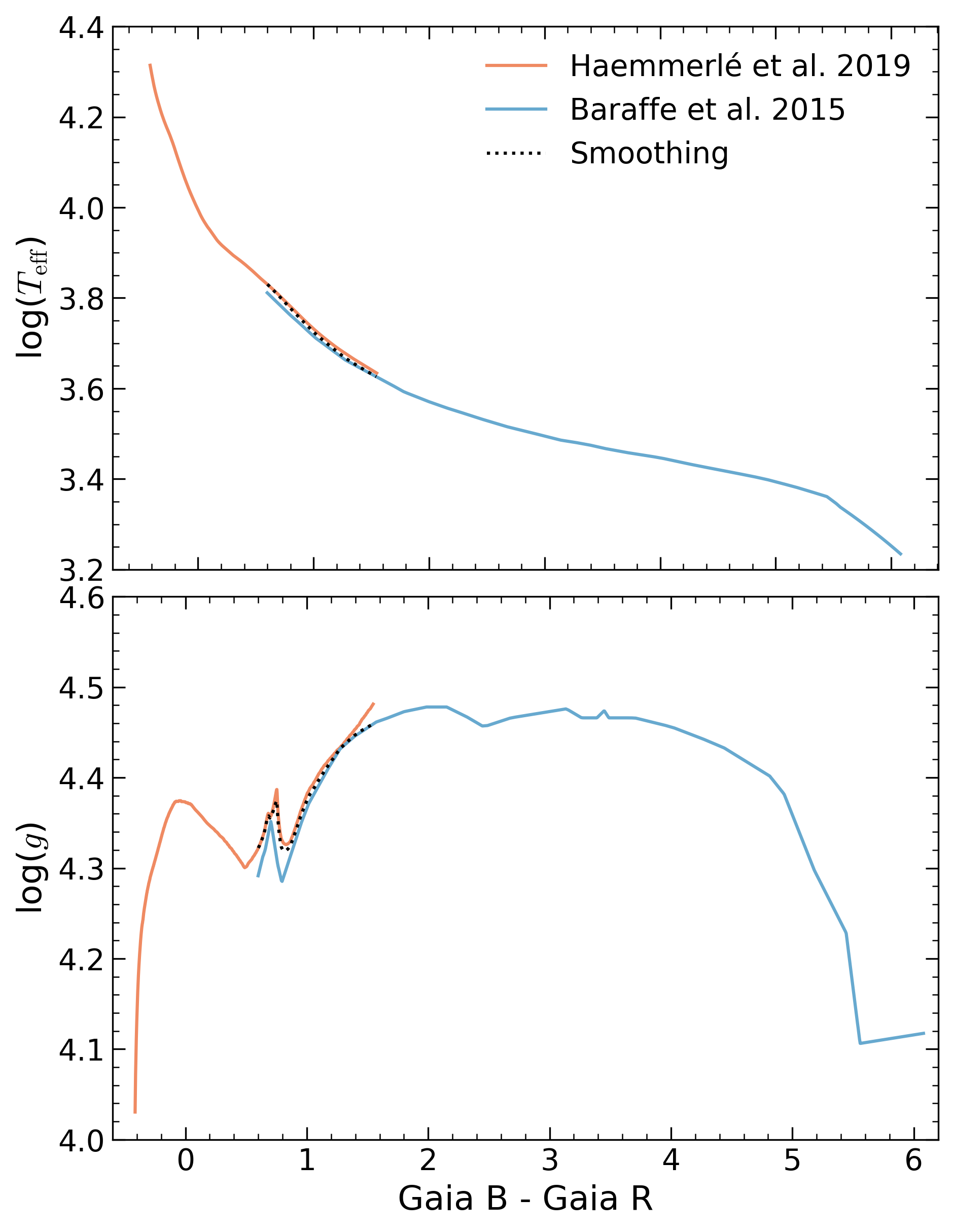}
    \caption[24 Myr isochrones for the Gaia B$-$R colour]{24 Myr isochrones of log($T_\textrm{eff}$) and log($g$) versus \textit{Gaia} $B-R$ colour corresponding to $\beta$PMG from the \citep{Bara15} (blue) and \citep{Haem19} (orange) evolutionary models. Similar curves for TWA are not displayed for clarity, but exhibit very similar variations. The black dotted lines indicate the smoothing of the two models in their overlapping region.} 
    \label{fig:isochrones}
\end{figure}

Prior to performing imaging simulations on each individual target in the sample it is necessary to generate their corresponding spectral energy distributions. This process begins by matching the \textit{Gaia B-R} colour \citep{Gaia16, Gaia18} of each target to a corresponding effective temperature ($T_{\mathrm{eff}}$) and log($g$) using theoretical stellar isochrones. The isochrones used are those from \citet{Bara15}, covering $0.07-1.4$~$M_\odot$, and \citet{Haem19} covering $0.8-120$~$M_\odot$. In order to retrieve as accurate a value as possible, each set of isochrones are interpolated to the age of the system of interest before the values of $T_{\mathrm{eff}}$ and log($g$) are determined. Additionally, for those objects that lie in the overlapping region between the two models we compute a weighting 
\begin{equation}
    \alpha = \frac{G_{B-R,\mathrm{target}}-\mathrm{max}(G_{B-R,\mathrm{Haemmerl\acute{e}}})}
    {\mathrm{min}(G_{B-R,\mathrm{Baraffe}})-\mathrm{max}(G_{B-R,\mathrm{Haemmerl\acute{e}}})}
\end{equation}
where $G_{B-R,x}$ is the \textit{Gaia B-R} colour for a target or model $x$. This weighting is then used to compute the values of $T_{\mathrm{eff}}$ and log($g$) for these objects using the linear relation
\begin{equation}
    Q = \alpha Q_\mathrm{Baraffe} + (1-\alpha)Q_\mathrm{Haemmerl\acute{e}}
\end{equation}
where $Q$ is the value of interest, and $Q_\mathrm{Baraffe}$ and $Q_\mathrm{Haemmerl\acute{e}}$ are the corresponding model values given an initial $G_{B-R,\mathrm{target}}$ value. Examples of these isochrones for $\beta$PMG, including the linear relation to smooth the overlapping region, are shown in \fref{fig:isochrones}.

SEDs for each object are then found by matching the determined $T_{\mathrm{eff}}$ and log($g$) values to theoretical spectra. As \citet{Haem19} do not provide spectra corresponding to their evolutionary models, for any objects with temperatures above the $\sim$7000~K limit of the \citet{Bara15} models we instead use the BOSZ models of \citet{Bohl17} (see also \citealt{Mesz12}). In all cases we assume solar metallicity during the spectral selection. Finally, each SED is normalised to its respective target's magnitude in the \textit{WISE W2} bandpass \citep{Wrig10, Cutr13}.

%%%%%%%%%%%%%%%%%%%%%%%%%%%%%%%%%%%%%%%%%%%%%%%%%%%%
\subsection{\texttt{PanCAKE} Simulations}\label{sec:pcake}
\begin{figure}
    \centering
    \includegraphics[width=\columnwidth]{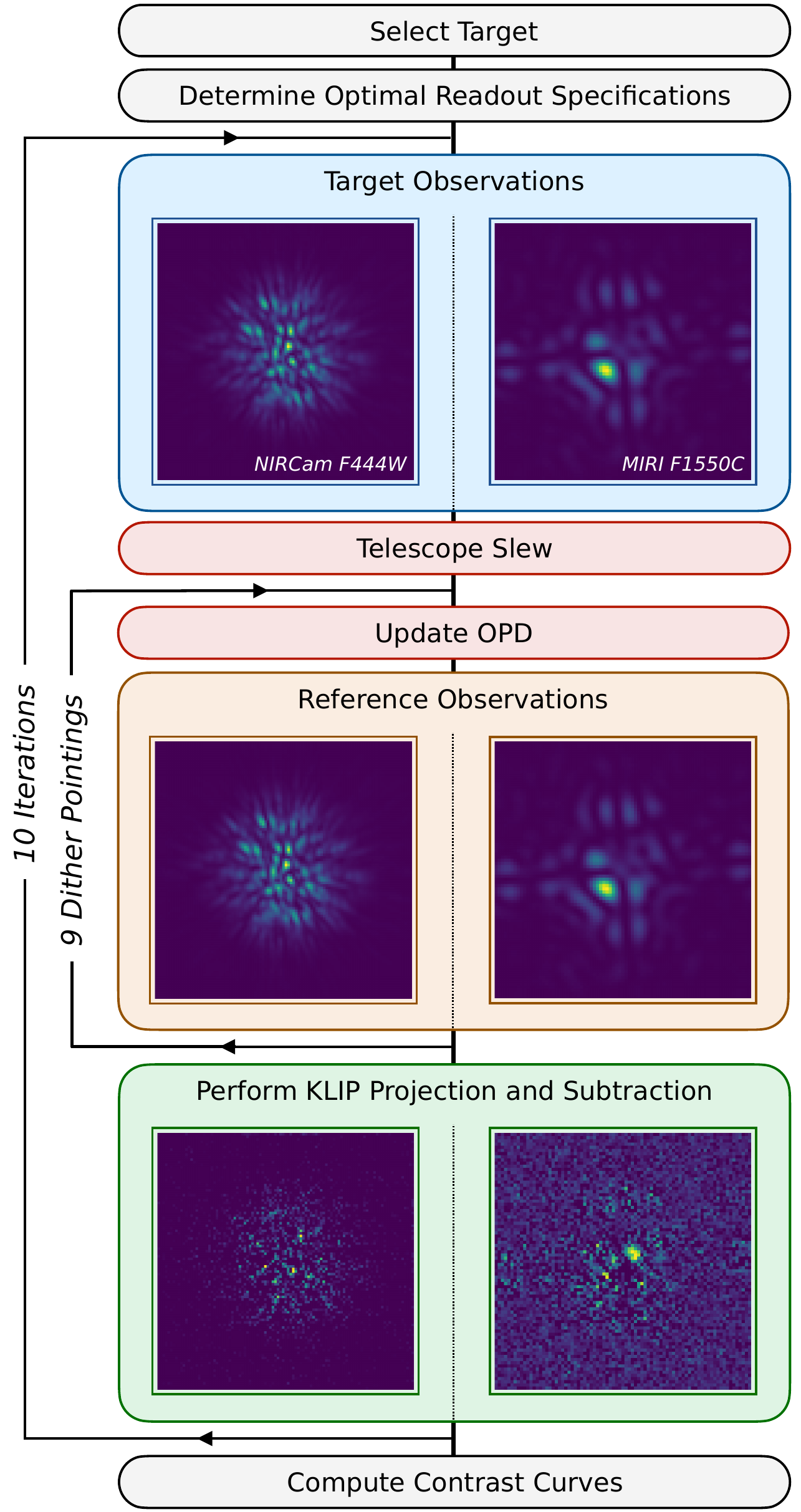}
    \caption{Block diagram representation of the simulation process. Example target, reference, and subtracted images are displayed for one of the sample targets in both the F444W (\textit{Left}) and F1550C (\textit{Right}) filters. In all images, any observed structure is a direct result of the residual scattered starlight or fundamental limitations in the subtraction process, and not descriptive of the spatial profile of a true astronomical object. The subtracted images are displayed on an intensity scale 100 times smaller than that of the corresponding target and reference images.} 
    \label{fig:block_diagram}
\end{figure}

All simulations are performed using the {\sc python} package \texttt{PanCAKE}\footnote{Pandeia Coronagraphy Advanced Kit for Extractions; \href{https://github.com/spacetelescope/pandeia-coronagraphy}{https://github.com/spacetelescope/pandeia-coronagraphy}} \citep{VanG16,Perr18, Gira18}, which is based on the official \textit{JWST} exposure time calculator \texttt{Pandeia} \citep{Pont16}. Given a desired input scene, \texttt{PanCAKE} is capable of producing corresponding 2D simulated images for all coronagraphic observations with the NIRCam and MIRI instruments aboard \textit{JWST}. For every object in the sample we simulate observations using NIRCam's MASK335R with the F356W and F444W filters, and MIRI's F1140C and F1550C masks. A block diagram demonstrating the primary steps performed in throughout the simulation process is displayed in \fref{fig:block_diagram}, alongside example simulated images in both the F444W and F1550C filters. Any observed structure in these images is a result of the residual scattered starlight or fundamental limitations in the subtraction process, and not descriptive of the spatial profile of a true astronomical object. Particularly noteworthy is a bright speckle in the lower left quadrant of the displayed F1550C target and reference images, which is in fact ubiquitous to all of the performed MIRI simulations. This particular feature is driven by the default initial optical telescope element wavefront map of \texttt{PanCAKE}, however upon testing all nine other available initial maps, which produce different residuals, we find that the final determined contrast limits are largely unaffected and this choice of default map does not impact our results. 

\subsubsection{Target and Reference Observations}
For each target, simulated observations are performed on the target star and a reference assumed to be identical in spectral type and magnitude to that of the target. Whilst mismatches in spectral type can lead to non-optimal PSF subtraction, differences on the order of a few subtypes are unlikely to significantly impact the estimated contrast performance \citep{Perr18} and we therefore assume that a viable reference can be found for every target in the sample. Unlike the target observations, we repeat the reference star observations nine times following a circular small-grid dither (SGD) pattern. Despite the time intensive nature of such a procedure, utilising a SGD technique significantly improves the contrast performance of \textit{JWST} observations \citep{Soum14, Lajo16} and as such will likely be necessary to reach the true contrast limits for \textit{JWST} coronagraphy. Furthermore, the expensive nature of these observations may be mitigated somewhat during a true observation by selecting a reference star brighter than that of the target. 

In reality it is not possible to centre the target or reference behind the coronagraphic mask due to the intrinsic pointing accuracy of \textit{JWST}. We include this effect in the simulations by applying a target acquisition error equal to a random draw from a normal distribution with a standard deviation equal to 5 mas. However, these offsets between target and coronagraphic mask lead to unique variations in the resultant simulated image and therefore biases in the determination of the simulated contrast curve. In these simulations we mitigate such affects by repeating each individual simulation 10 times in order to generate a statistical sample from which the contrast curve can later be determined.

When performing a default simulation, the \texttt{PanCAKE} package utilises a library of PSFs which are precomputed across the coronagraphic field of view. As a result, variations in the simulated images due to small offsets such as the target acquisition error will not be accurately represented. We remedy this issue in our simulations by enabling the \texttt{on\char`_the\char`_fly\char`_PSFs} setting within \texttt{PanCAKE}. This setting circumvents the use of the precomputed library and instead calculates the precise PSF using the \texttt{WebbPSF} dependency \citep{Perr14}. In order to make the simulations computationally tractable we reduce the wavelength sampling from the default of 150, to 41 through the \texttt{wave\char`_sampling} setting. Despite this reduction, variations from the true PSF are on the order of $<$1\% (Marshall Perrin, private communication).  

For both the NIRCam and MIRI simulated observations an exposure time of $\sim$3600~s was selected. The exact choice of exposure time most significantly impacts the sensitivity estimations at the widest angular separations, where the contrast is relatively unchanged with separation and is primarily limited by background and photon noise. However, at shorter angular separations where the contrast is a steep function of separation, the exposure time has a reduced impact as it is the residual stellar noise that primarily limits the achievable sensitivity. The chosen exposure time has the advantage of being short enough to be observationally feasible, yet long enough that we observe no appreciable improvement to the simulated contrast at these shorter separations with an even longer exposure. For the vast majority of the simulated targets the angular separation at which the sensitivity transitions to the background and photon noise limited regime corresponds to a physical separation of $\gtrsim$100~au. At such separations the occurrence rate of planetary-mass objects is greatly reduced as evidenced by observational studies \citep{Durk16, Brya16, Viga17, Niel19, Baro19, Viga20}, and both core accretion and gravitational instability population synthesis models \citep{Ida04, Forg13, Forg15, Emse20}. Therefore, whilst the mass sensitivity as a function of exposure time is not explicitly investigated, the simulations shown in this study represent an estimate of the practicable sensitivity limits of \textit{JWST} coronagraphy at separations where planetary-mass objects are predicted to be the most common. We note that the observational studies mentioned have at best been sensitive to objects $\gtrsim$1~$M_\textrm{J}$, and therefore the frequency of sub-Jupiter mass objects at these separations has still not been explicitly constrained. If in actuality a population of such objects exists that is enhanced relative to predictions from population synthesis models, it will be worthwhile to revisit the presented sensitivity limits in Section \ref{sec:discussion} to account for potential improvements in the background dominated regime with an increased observation time, however we elect not to perform such an analysis in this work. Above all, we stress that for an actual observation the duration should be carefully selected in order to reach a desired sensitivity limit for a specific target and separation of interest, especially for angular separations in the background and photon noise limited regime. 

\subsubsection{Readout Specifications}
The NIRCam detectors host a broad range of readout modes, with nine distinct readout patterns and up to either 10 or 20 groups per integration. Whilst this provides significant versatility, finding the optimal readout mode for a desired observation can be a non-trivial task. Although it is technically possible to perform the simulation for every possible readout mode to assess which performed the best, this is very computationally intensive and hence not practical. Alternatively, we estimate the optimal readout mode using the \texttt{ramp\_optimize} function of the \texttt{pyNRC} software package (Leisenring et. al, in preparation\footnote{\href{https://pynrc.readthedocs.io/en/latest/}{https://pynrc.readthedocs.io/en/latest/}}). Given a NIRCam observational setup and a desired integration time, this function can quickly estimate the achievable SNR of a target object for each possible readout mode. We use this function to select the input readout mode for every \texttt{PanCAKE} simulated target by finding the corresponding \texttt{ramp\_optimize} mode which: provides the maximal SNR of a synthetic companion object 20 magnitudes fainter than the target star, has a total integration time less than 3600~s, and does not saturate the detector. The choice of such a faint companion is made as this preferentially selects readout modes best for observing objects with the largest contrast to their host stars.

For the MIRI simulations we instead perform a custom optimisation to determine the optimal readout parameter for every target in the sample. The MIRI detector only has two possible readout patterns~$-$~SLOW or FAST~$-$~and the SLOW pattern is only recommended for parallel observations where the overall data volume in the FAST pattern would be too high. Therefore only the number of groups (NG) and integrations (NI) needs to be optimised for the simulations (see \citet{Ress15} for further discussion on MIRI readout specifications). For each target object we begin by determining the fraction of detector saturation for a desired MIRI filter using a simplistic \texttt{PanCAKE} simulation with a FAST readout pattern, NG=5, and NI=1. This fraction is then inverted to determine the maximal NG without going beyond a saturation fraction of 80\%. For maximal NG values $\leq$100 we determine the maximal NI that corresponds to a total observation time less than 3600~s and adopt these maximal values as the input values of the simulation. The reason for such a cutoff is to avoid non-ideal detector effects which are more pronounced for shorter integrations \citep{Ress15}. For maximal NG values $>$100 we instead adopt the highest NG value possible that: a) results in individual integration times less than 280~s, and b) has a corresponding NI value that results in a total observation time between 3580 and 3600~s. The first restriction will be necessary for true on-sky observations as long integrations will be more significantly affected by cosmic rays, with every pixel being affected, either directly or indirectly, by $\sim$1000~s\footnote{\href{https://jwst-docs.stsci.edu/mid-infrared-instrument/miri-observing-strategies/miri-cross-mode-recommended-strategies}{https://jwst-docs.stsci.edu/mid-infrared-instrument/miri-observing-strategies/miri-cross-mode-recommended-strategies}}. The second restriction ensures that all of the observations have approximately similar integration times.

\subsubsection{Accounting for Thermal Drifts}
The slew performed by \textit{JWST} when moving from target to reference star will inherently cause a variation in the observatory pitch angle relative to the Sun. This variation leads to a difference in the overall illumination of the observatory between observations and will induce a thermally driven wavefront drift which is currently expected to be the primary driver of variations in \textit{JWST} optical telescope element \citep{Perr18}. Any such variations will inhibit the ability to perform an accurate PSF correction and therefore will affect the achievable contrast. 

We include the effects of thermal drifts throughout the simulated observations using the \texttt{thermal\_slew} function included in the \texttt{webbpsf} {\sc python} software package \citep{Perr14}. By providing a desired slew start pitch angle, end pitch angle, and the elapsed time, this function can model the variation in the optical path difference (OPD) map of each primary mirror segment which in turn can be provided to the \texttt{PanCAKE} simulation. During these simulations the OPD map is updated after the target observation and then after every consecutive dither for the reference star observation. 

\begin{figure}
    \centering
    \includegraphics[width=\columnwidth]{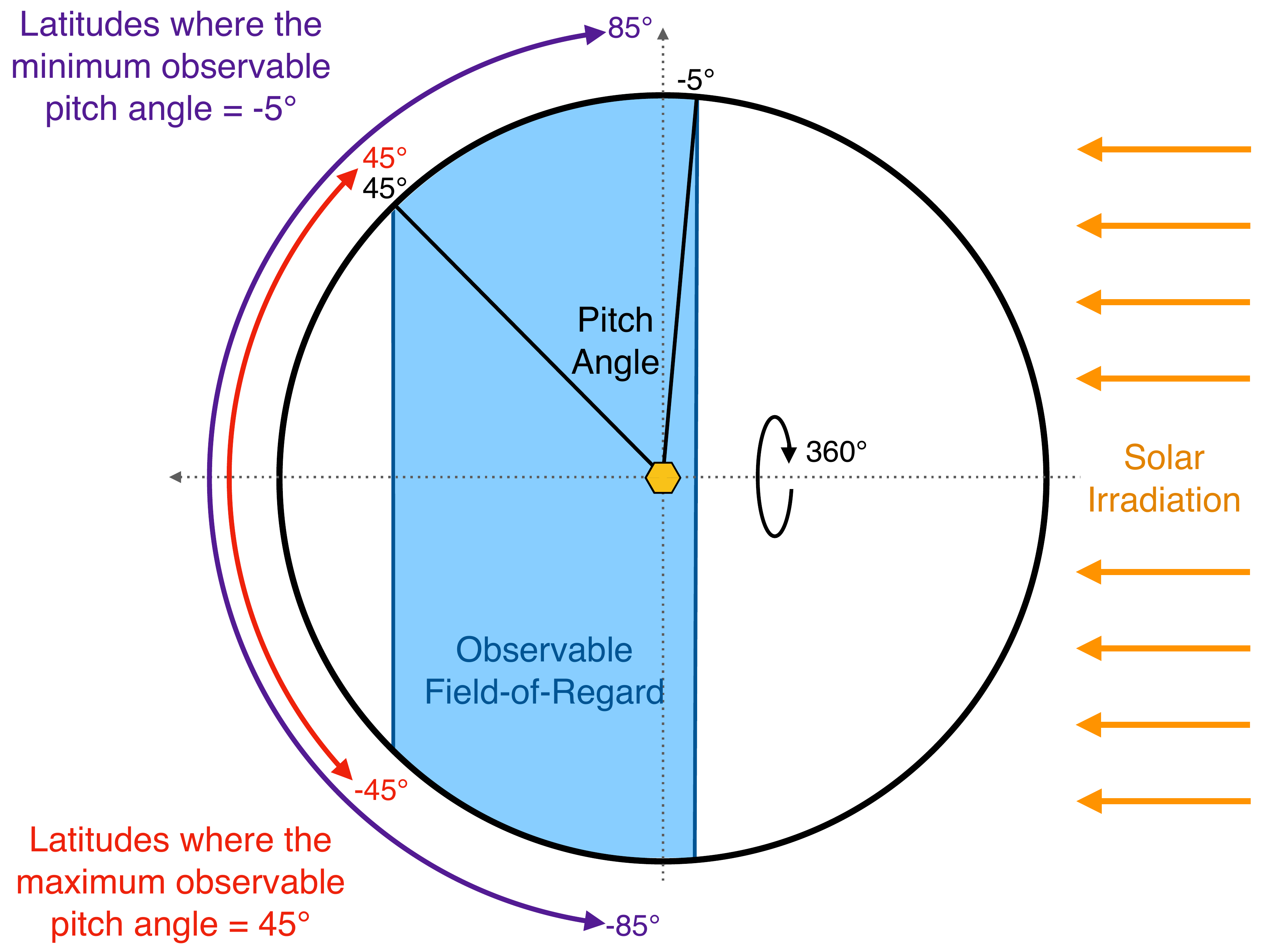}
    \caption{A 2-dimensional representation of the sky-projected observable field-of-regard (shaded region) of \textit{JWST} (central hexagon) at any given time. In reality the observable field-of-regard moves as \textit{JWST} orbits the Sun, and covers the entire sky every year. As the pitch angle is constrained between 45$^\circ$ and $-5^\circ$, only ecliptic latitudes between 45$^\circ$ and $-$45$^\circ$ can be observed at all possible pitch angles during the course of a year. Perpendicular to the pitch angle axis, \textit{JWST} is able to rotate a full 360$^\circ$. } 
    \label{fig:fov_diagram}
\end{figure}

Selecting a slew start pitch angle is not straightforward, as the observatory pitch angle can vary between 45$^\circ$ and $-5^\circ$ as shown in \fref{fig:fov_diagram}. Furthermore, the minimum and maximum observable pitch angle are dependent on the target ecliptic latitude, where only objects at latitudes between $-45^\circ$ and $45^\circ$ can be observed across the full pitch angle range. For each independent object in the sample a corresponding minimum and maximum pitch angle is therefore calculated as $\max(-5,\lvert\beta_e\rvert-90)$ and $\min(45,90-\lvert\beta_e\rvert)$ respectively, where $\beta_e$ is the ecliptic longitude of the object. We assume average yearly pitch angles for the slew start angles by taking the midway point between these minimum and maximum observable pitch angles. 

To select the slew end angle we must first assume an offset angular distance between the target and reference star. As we do not explicitly select real reference stars for each of the targets it is necessary to instead determine a ``typical'' angular offset. To approximate this for the simulations we take the average offset for all \textit{JWST} Early Release Science (ERS) and Guaranteed Time Observer (GTO) coronagraphic observations, resulting in a value of $\sim$6.6$^\circ$. However, given that the observatory can rotate a full 360$^\circ$ in the axis perpendicular to the pitch axis (see \fref{fig:fov_diagram}), the average slew offset does not directly correspond to an average change in pitch angle between target and reference. Neglecting spherical geometric effects, for a given slew offset the change in pitch angle, $\delta_{\textrm{p}}=\delta_{\textrm{s}}\sin(\phi)$, where $\delta_{\textrm{s}}$ is the slew offset and $\phi$ is the angle between the direction of the slew offset and the direction of the pitch offset. Whilst the value of $\phi$ can vary freely from 0$-$2$\pi$, the absolute average change in the pitch angle can be calculated by determining the arithmetic mean of $\delta_{\textrm{P}}$ across a circle quadrant of radius $\delta_{\textrm{s}}$ from 0$-\pi/2$,
\begin{equation}
\delta_{\textrm{p,av}} = \frac{2\delta_{\textrm{s}}}{\pi} \int^{\pi/2}_{0} \sin(\phi) d\phi,
\end{equation}
which for the assumed average slew offset of 6.6$^\circ$ is equal to 4.2$^\circ$, and is treated as the difference between the slew start and end pitch angles for all the sample objects. Whilst it is possible to add or subtract this value from the slew start pitch angle to obtain a slew end pitch angle, we do not observe any significant differences in the final contrast curve and as such opt to add it. 

To determine the elapsed time between the slew start pitch angle and the slew end pitch angle we refer to the official \textit{JWST} slew times\footnote{\href{https://jwst-docs.stsci.edu/jppom/visit-overheads-timing-model/slew-times}{https://jwst-docs.stsci.edu/jppom/visit-overheads-timing-model/slew-times}}. An overall slew distance of 6.6$^\circ$ corresponds to a total slew time of $\sim$980~s, including 284~s for a necessary guide star reacquisition. In the simulations we adopt a slightly more conservative value of 1000~s for the elapsed time between the end of the target observation and the start of the reference observation to account for target acquisition procedures. Finally, after each subsequent dithered reference observation, we recompute the OPD map by adding one hour to the elapsed time. 

\subsubsection{Contrast Curve Determination}\label{sec:ccurves}
Following the generation of all of the 2D simulated images, corresponding contrast curves are computed which describe the limiting sensitivity of the synthetic observations. As 10 realisations of simulated target and reference images were produced for each object in the sample, it is first necessary to subtract a stellar PSF from each of the target images. This subtraction is performed for each target image realisation using a synthetic PSF generated from its corresponding reference star images following the KLIP algorithm of \citet{Soum12}, as implemented by the \texttt{klip\char`_projection} function within \texttt{PanCAKE}. An estimation of the radial contrast curve for each target is then calculated using the default \texttt{Pandeia} correlation matrix method on the remaining ensembles of subtracted images as described in Appendix \ref{corrmatrix_snr}. In all cases we exclude separations shorter than 1~$\lambda/D$, where $\lambda$ is the central wavelength of the used filter and $D$ is the \textit{JWST} primary mirror diameter of 6.5~m, as objects at these separations will be indistinguishable from the central stellar emission. 

To reduce computational intensity the observed field of view in \texttt{PanCAKE} is reduced to 6.3$\arcsec\times$6.3$\arcsec$ and 8.8$\arcsec\times$8.8$\arcsec$ from 20$\arcsec\times$20$\arcsec$ and 24$\arcsec\times$24$\arcsec$ for NIRCam and MIRI respectively. Additionally, the estimated contrast at the widest simulated separations is imprecise as it is calculated from only a few pixels in the image. To alleviate these effects we extend each NIRCam and MIRI contrast curve to 10$\arcsec$ and 12$\arcsec$ respectively, assuming that the contrast at separations beyond 90\% of the widest simulated separation is constant and equal to the contrast at 90\% of the widest simulated separation. Whilst larger radial separations are possible, exoplanets at these separations would not be observable at all roll angles. In order to avoid any overestimation of the detection probability maps determined in Section \ref{sec:results} we do not include these widest separations in our analysis. Each contrast curve is then divided by the respective NIRCam\footnote{\href{https://jwst-docs.stsci.edu/near-infrared-camera/nircam-instrumentation/nircam-coronagraphic-occulting-masks-and-lyot-stops}{https://jwst-docs.stsci.edu/near-infrared-camera/nircam-instrumentation/nircam-coronagraphic-occulting-masks-and-lyot-stops}} or MIRI \citep{Dani18} coronagraphic transmission profile to incorporate the intrinsic IWA restrictions of these observations. Finally, all contrast curves are converted from angular separation to physical separation using the reported \textit{Gaia} distances \citep{Gaia16, Gaia18} for each individual target.

At this stage it is common to calculate a final contrast curve for a given observation in terms of an integer multiplication, $n$, of the base contrast curve, representing the threshold at which at object would be $n$ times brighter than the noise, $\sigma$. However, such a measure fails to account for the limitations of the small number statistics at the innermost separations \citep{Mawe14}, and irrespective of the value of $n$, corresponds to a fraction of true positive detections of only 50\% \citep{Jens18}. To account for these effects, we perform a final correction to the more traditional 5$\sigma$ simulated contrast curves following the prescription detailed in \citet{Ruan17, Ston18}, using a true positive detection fraction of 95\% and a total of 0.01 false detections per image. As the nature of this correction is intrinsically linked to the number of available resolution elements, which is a function of observation wavelength, it is unique to each simulated filter. In general, these corrections result in variations to the sensitivity limits on the order of $-0.3$ magnitudes at the widest separations, and up to $\sim$1 magnitude at the innermost separations.

\begin{figure}
    \centering
    \includegraphics[width=\columnwidth]{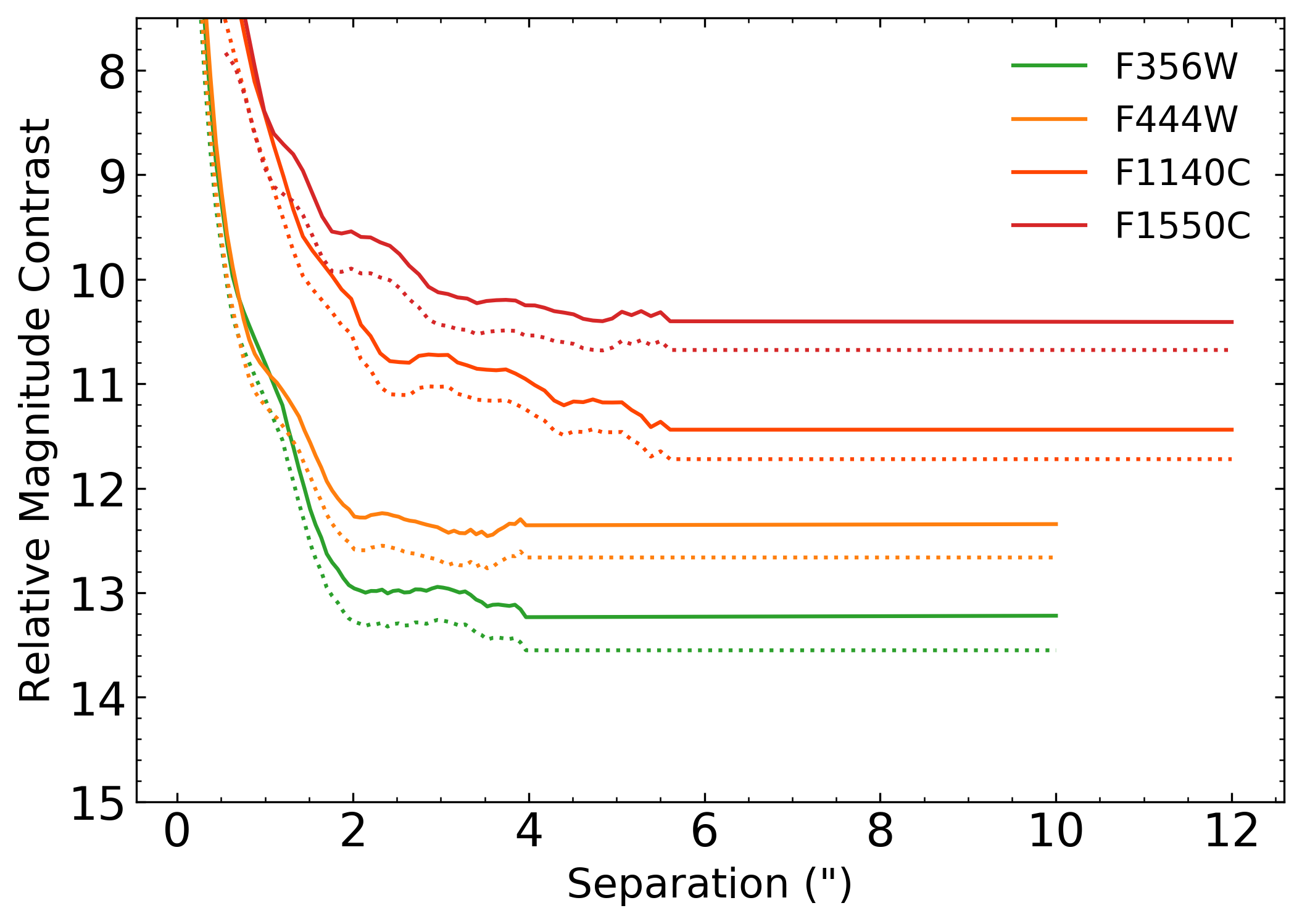}
    \caption[Simulated JWST NIRCam and MIRI contrast curves]{From bottom to top, example 95\% completeness (solid) and classical 5$\sigma$ (dotted) simulated contrast curves for the F356W, F444W, F1140C, and F1550C filters for a single target within the sample. Contrasts at separations beyond the simulated field of view, yet still within the observationally possible field of view, are assumed to be constant and equal to the contrast at 90\% of the widest simulated separation.}
    \label{fig:individual_concurves_v2}
\end{figure}

An example set of these 95\% completeness contrast curves are plotted alongside the more traditional 5$\sigma$ contrast curves for a single object within the sample in \fref{fig:individual_concurves_v2}. In all of the simulations the NIRCam observations reach a superior contrast than the MIRI observations. However, as the SED of both a star and planet vary as a function of wavelength, the relative magnitude contrast between them will also vary. To assess which filter is best suited towards detecting the lowest mass exoplanets, it is beneficial to present these limits in terms of mass.

%%%%%%%%%%%%%%%%%%%%%%%%%%%%%%%%%%%%%%%%%%%%%%%%%%%%%%%%%%%%%%%%%%%%%%%%%%%%%%%%%%%%%%%%%%%%%%%%%%%%%%%%%
%%% RESULTS
%%%%%%%%%%%%%%%%%%%%%%%%%%%%%%%%%%%%%%%%%%%%%%%%%%%%%%%%%%%%%%%%%%%%%%%%%%%%%%%%%%%%%%%%%%%%%%%%%%%%%%%%%
\section{Detection Probability Modelling}\label{sec:results}
\subsection{Mass Sensitivity Estimation: Model Selection} 
To estimate the detectable mass limits of \textit{JWST} coronagraphy it is necessary to convert the determined contrast curves from the standard representation in terms of magnitude, to one in terms of mass. To do so we make use of planetary evolution models, which predict the magnitude of an object within a filter of interest given its mass. Specifically, we rely on the latest models of low mass planets spanning $\sim5$~$M_\textrm{E}-2$~$M_\textrm{J}$ from \citet{Lind19} (\texttt{BEX}), and more massive giant planets and brown dwarfs spanning $\sim$0.5~$M_\textrm{J}-75$~$M_\textrm{J}$ from \citet{Phil20} (\texttt{ATMO}). Whilst such a conversion is necessary for this work, it is not without its limitations. Chief of which is that in reality there is no single relation between mass and luminosity for a given object at a given time, due to the variety of distinct formation and evolution pathways it may have undergone \citep{Mord17, Emse20}. This effect is most prominent at younger ages (>3~Myr, \citealt{Mord17}), and for an object at 20~Myr corresponds to a maximum mass uncertainty of $\sim$1~$M_\textrm{J}$ for masses below 10~$M_\textrm{J}$ \citep{Emse20}. We do not attempt to account for such effects in the performed mass conversions, but instead emphasise that the resulting sensitivity estimations are better described as average mass sensitivities to the entire planetary population, and not an individual object. Another limitation to the implementation of these models is that current predictions for the initial entropy conditions of planetary mass objects are uncertain \citep{Marl07, Spie12, Marl14}. The used \texttt{ATMO} models are based on a hot-start formation and at the adopted ages of TWA and $\beta$PMG the \texttt{BEX} models also closely match a hot-start formation (Fig. 9 in \citealt{Lind19}). In a contrasting cold-start scenario an object of a given mass will have a lower luminosity, and will therefore be more difficult to detect. For a 1~$M_\textrm{J}$ object observed in the near- to mid-infrared at 10~Myr this difference can be as significant as a magnitude in brightness \citep{Spie12}. Such a discrepancy will negatively impact the presented mass sensitivities, however we do not explicitly investigate this effect.

\texttt{ATMO} offers three different sets of evolutionary models: one at chemical equilibrium, and the other two at chemical disequilibrium assuming different strengths of vertical mixing. Given that all of the \texttt{BEX} models are computed at chemical equilibrium, we do not explore the effects of disequilibrium chemistry on \textit{JWST} sensitivity limits as part of this study and select the equilibrium models at solar metallicity. 

Similarly to \texttt{ATMO}, \texttt{BEX} offers a range of different sets of evolutionary models based on different atmospheric models: one using the \texttt{Ames-COND} grid \citep{Alla01}, 14 based on the \texttt{petitCODE} grid \citep{Moll15, Saml17}, and one using the \texttt{HELIOS} grid \citep{Mali17}. As \texttt{petitCODE} has recently been benchmarked against \texttt{ATMO} \citep{Baud17}, we select these models for this study. Of the 14 evolutionary model sets produced from the \texttt{petitCODE} grid, many incorporate varying levels of metallicity or clouds. Only Na$_2$S and KCl clouds are included in these models and therefore water clouds, which are expected impact the spectra of objects with temperatures $\lesssim$400~K \citep{Morl14}, are neglected. Furthermore, recent work has suggested that the high nucleation energy barrier of Na$_2$S strongly inhibits its formation and therefore its inclusion as a dominant cloud species may not be strictly accurate \citep{Gao20}. For these reasons, and as none of the \texttt{ATMO} models of \citet{Phil20} include the effects of cloud opacity, we select the solar metallicity \texttt{petitCODE} models without any cloud opacity and retain model consistency between mass ranges. 

As we are unable to simultaneously include both cloud formation and disequilibrium chemistry processes with the chosen evolutionary models, it is not straightforward to determine exactly how their complex interplay would affect the overall mass sensitivity calculations presented in this work. However, as these processes play a significant role in the overall structure and composition of sub-stellar atmospheres, differences will likely exist. 

The presence of a silicate or alkali cloud deck acts to limit the atmospheric depth from which flux can readily emerge, primarily resulting in reductions of the $\sim$1$-$2~$\mu$m emission for objects with temperatures <1300 K. This flux is then redistributed, producing an opposing increase in emission at wavelengths beyond $\sim$2~$\mu$m \citep{Morl12, Char18}. However, for objects at temperatures below 400~K, which represent our best case limiting mass sensitivity, this increase is negligible \citep{Morl12}. In contrast, the formation of water clouds below $\sim$400~K can substantially affect the emitted flux at infrared wavelengths. In the case of a large fractional cloud coverage of 80\% at a 200~K effective temperature, the emitted flux could be reduced by approximately half an order of magnitude at ~$\sim$4.5~$\mu$m, and increased by approximately two orders of magnitude at ~$\sim$3.5~$\mu$m \citep{Morl14}. Nevertheless, these effects become less significant with increasing temperature, or decreasing cloud coverage. Whilst we do not aim to specifically quantify the overall impact of clouds on our simulations, in a qualitative sense, and in this respect alone, the estimates of the limiting mass sensitivity in the F356W and F444W filters may therefore be considered somewhat pessimistic or optimistic respectively. 

Disequilibrium chemistry is primarily considered through enrichment of molecular species in the upper atmosphere through upwards vertical mixing from deeper and hotter regions of the atmosphere. Recently, such an effect has been inferred ubiquitously in a sample of the coolest brown dwarf atmospheres, ranging from $250-750$~K in effective temperature, through the enhancement of CO absorption from $\sim$4.5$-$5.0~$\mu$m in their \textit{M}-band spectra \citep{Mile20}. Among these brown dwarfs the scale of the absorption varies from object to object, although in general it results in reductions in the emitted flux by approximately a factor of two compared to an equilibrium model. However, CO is likely not the only species that may be enhanced through disequilibrium chemistry, and species such as CO$_2$, HCN, C$_2$H$_2$, PH$_3$, and GeH$_4$ may also reduce the emitted flux at infrared wavelengths \citep{Morl18}. Unfortunately, observations of disequilibrium chemistry in exoplanetary atmospheres are still relatively sparse, and its significance and nature as a function of planetary-mass and temperature is still not well understood. We therefore do not attempt to quantify exactly how disequilibrium processes may affect our simulations either. Nonetheless, in the absence of disequilibrium enhanced absorption it can be assumed that the limiting mass sensitivities for all of the simulated filters are likely more optimistic. 

Quantitatively assessing the interplay between model predictions of \textit{JWST} performance, planetary evolution, and planetary atmospheres to estimate the limiting mass sensitivity of \textit{JWST} is a complex and constantly evolving task. As such, the mass sensitivity estimates presented in this work must be considered in context of the underlying models we have selected. Understanding and describing the exact impact of many tunable properties within this process $-$ such as the presence of clouds and disequilibrium chemistry as described above $-$ is currently difficult to accomplish due to the limited to non-existent nature of observations that inform them.  As these models are improved and updated and such properties are better understood (perhaps even in response to \textit{JWST} observations themselves), it will be advantageous to reexamine the determined limiting mass sensitivities and identify any significant improvement or decline in performance. 

\subsection{Mass Sensitivity Estimation: Model Application} 
In the case of \texttt{ATMO}, evolutionary models have already been computed for all \textit{JWST} coronagraphic filters, including those used as part of this study. However, the \texttt{BEX} models have only been computed for a subset of the \textit{JWST} photometric filters. To produce new \texttt{BEX} evolutionary magnitude tracks we first calculate the synthetic magnitude in all the coronagraphic filters used throughout this study for each of the chosen \texttt{petitCODE} models. The full throughputs used in this process are equal to the PCEs as calculated from the \textit{JWST} exposure time calculator \texttt{Pandeia}, and are displayed in \fref{fig:all_pces}. For each filter, we produce a corresponding 2-dimensional interpolation over these magnitudes in $T_\textrm{eff}-$log($g$) space. For each mass division in the existing \texttt{BEX} evolutionary tracks, we obtain the corresponding value of $T_\textrm{eff}$ and log($g$) and then pass these values to the previously described interpolations to determine the corresponding magnitudes for all filters used in this study at these mass divisions. To verify this method we also compute these magnitudes for the already calculated F356W photometric filter shown in \citet{Lind19}. The resulting differences between our calculation and that of \citet{Lind19} are <0.07~mag in all cases and are likely a result of different interpolation methods, or even the precision of the astronomical constants used in the underlying calculations. 

As both of these evolutionary models are computed across a range of specific ages, all of the following analyses were performed using interpolations to these models at the nominal ages of $\beta$PMG and TWA (see Section \ref{sec:ymg_selection}). First, the apparent magnitude of each target star in each of the used coronagraphic filters is added to its corresponding contrast curve as produced in Section \ref{sec:ccurves}. These magnitudes are computed using the SEDs generated in Section \ref{sec:sed_select} and the PCEs as shown in \fref{fig:all_pces}, and are displayed in Tables \ref{tab:bpmg_sample} and \ref{tab:twa_sample}. In essence, this process converts the contrast curves from a relative magnitude contrast, to an absolute detectable magnitude limit. This limit can then be converted to an absolute detectable mass limit using the interpolation between mass and magnitude in a desired filter from the aforementioned evolutionary models. 

\begin{figure}
    \centering
    \includegraphics[width=\columnwidth]{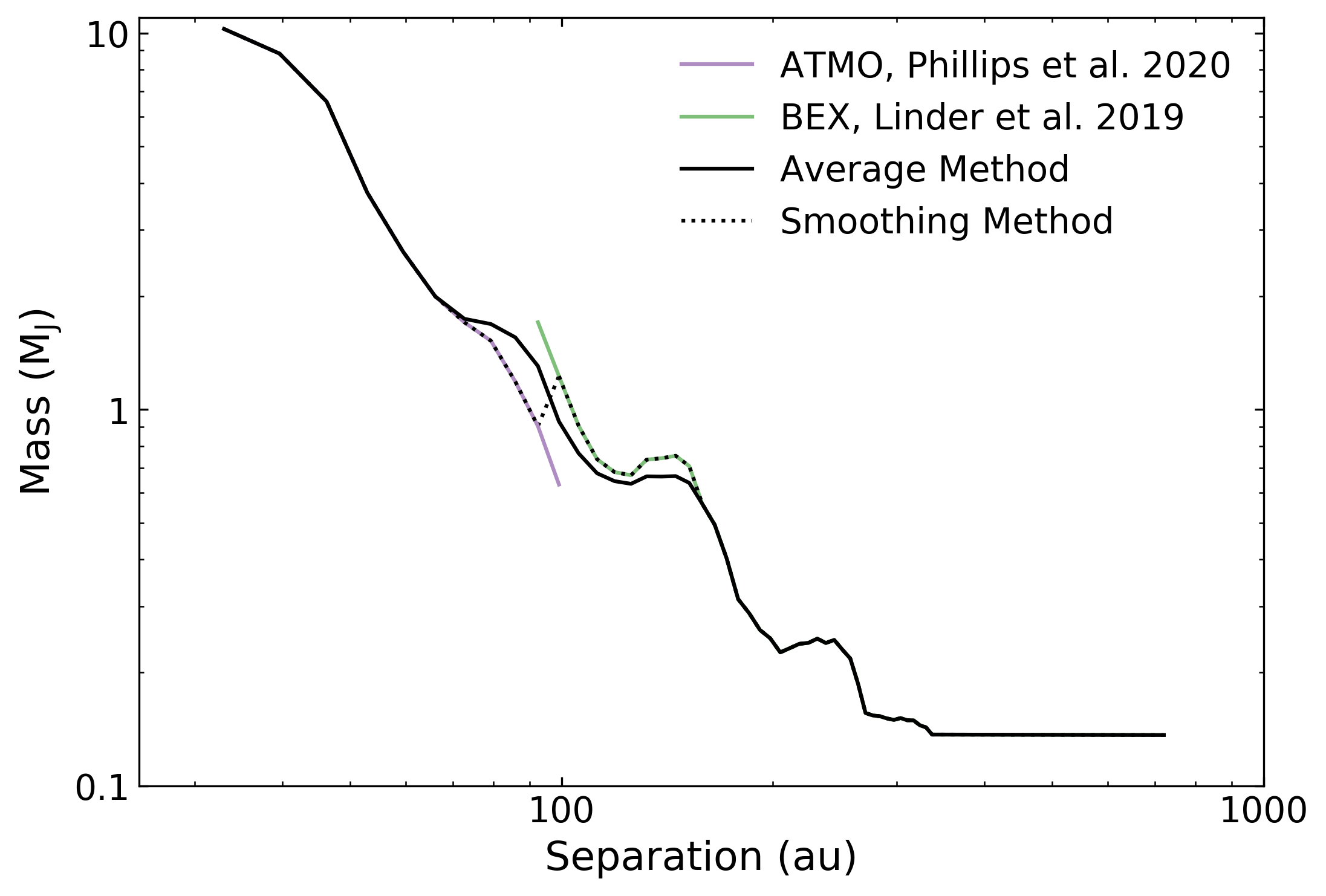}
    \caption[Averaging interpolations of evolutionary models]{Mass sensitivity curves for a single example object within the sample. The difference between the \texttt{ATMO} (purple) and \texttt{BEX} (green) models causes a sharp discontinuity using a smoothing procedure similar to that described in Section \ref{sec:sed_select} (dotted black), however when the described averaging method is used (solid black) this discontinuity is not produced.}
    \label{fig:mass_evomodels}
\end{figure}

Similarly to the stellar evolutionary models in Section \ref{sec:sed_select}, there is an overlap in mass between the the \texttt{ATMO} and \texttt{BEX} models from 0.5$-$2~$M_\textrm{J}$. However, for many of the simulations this overlap is too small to produce an effective smoothed model in a similar fashion to that shown in Section \ref{sec:sed_select}. To account for the overlapping region, we instead take an average between the \texttt{ATMO} and \texttt{BEX} interpolations as a function of semi-major axis. As the \texttt{ATMO} and \texttt{BEX} models are not perfectly congruent, the smoothing method from Section \ref{sec:sed_select} produces discontinuities in the mass sensitivity curves as some values of the \texttt{ATMO} mass interpolation that are not in the averaged region will lie below the maximum averaged value. Similarly, some values of the \texttt{BEX} mass interpolation that are not in the averaged region will lie above the minimum averaged value. Whilst we do not seek to assess and account for the discrepancy between these evolutionary models explicitly, it is desirable to remove these discontinuities when determining the final mass sensitivity limits. This is performed by averaging the \texttt{ATMO} values that lie outside the overlapping region and are lower than the maximum averaged mass within the overlapping region with a straight line connecting the maximum \texttt{BEX} value and the closest but higher \texttt{ATMO} value. In a corresponding fashion, the \texttt{BEX} values that lie outside the overlapping region and are above the minimum averaged mass within the overlapping region are averaged with a straight line connecting the minimum \texttt{ATMO} value and the closest but lower \texttt{BEX} value. An example mass sensitivity curve is displayed in \fref{fig:mass_evomodels} alongside the overall model resulting from both the smoothing procedure from Section \ref{sec:sed_select}, which produces a discontinuity, and the averaging procedure described here, which does not. We note that the discrepancy between the two evolutionary models for a given mass are often $\gtrsim$1~mag and therefore cannot result from the slight differences arising from the independent calculation of the \texttt{BEX} evolutionary tracks. In the event that a magnitude value is too small (i.e. too bright) to be interpolated by the \texttt{ATMO} evolutionary models, we simply set the mass value to 75~$M_\textrm{J}$. This does not have any effect on the overall mass sensitivity limits, which are at much lower masses. Finally, in the event that a magnitude value is too large (i.e. too faint) to be interpolated by the \texttt{BEX} evolutionary models, we set the mass value to the minimum calculated mass value of the contrast curve. 

%%%%%%%%%%%%%%%%%%%%%%%%%%%%%%%%%%%%%%%%%%%%%%%%%%%%
\subsection{\texttt{\textbf{Exo-DMC}} Detection Probabilities}
Following the calculation of mass sensitivity limits for each of the objects we estimate detection probability maps using the Exoplanet Detection Map Calculator (\texttt{Exo-DMC}\footnote{\href{https://ascl.net/2010.008}{https://ascl.net/2010.008}}, \citealt{Bona20}). This is the latest (and for the first time in {\sc python}) rendition of the existing {\texttt{MESS}} (Multi-purpose Exoplanet Simulation System, \citealt{Bona12}) code, which utilises a Monte Carlo method to perform statistical analyses of the results from direct imaging surveys. \texttt{Exo-DMC} combines the information on the target stars with the instrument detection limits to estimate the probability of detection of a given synthetic planet population, ultimately generating detection probability maps. Specifically, \texttt{Exo-DMC} produces a grid of masses and physical separations of synthetic companions for each star in the sample, then estimates the probability of detection at each grid point given the provided detection limits. In the case of direct imaging observations, such as those simulated in this work, the potential for each synthetic companion to lie outside the instrumental field of view is also accounted for by generating a set of uniformly distributed orbital parameters for each point in the grid. This addition allows for the estimation of the range of possible projected separations corresponding to each value of semi-major axis.

In a similar fashion to its predecessors, \texttt{Exo-DMC} allows for a high level of flexibility in terms of possible assumptions on the synthetic planet population to be used for the determination of the detection probability. However, in this case we use the default setup, which uses flat distributions in log space for both the mass and semi-major axis and a Gaussian eccentricity distribution with $\mu =0$ and $\sigma = 0.3$ (following the approach of \citealt{Hogg10}, see \citealt{Bona13} for details). The detection probability maps generated for this study range from $0.01-75$~$M_\textrm{J}$ and $1-1000$~au, with a resolution of 500 in each dimension. 

%%%%%%%%%%%%%%%%%%%%%%%%%%%%%%%%%%%%%%%%%%%%%%%%%%%%
% DISCUSSION
%%%%%%%%%%%%%%%%%%%%%%%%%%%%%%%%%%%%%%%%%%%%%%%%%%%%
\section{Discussion}\label{sec:discussion}
\begin{figure*}
    \centering
    \includegraphics[width=\textwidth]{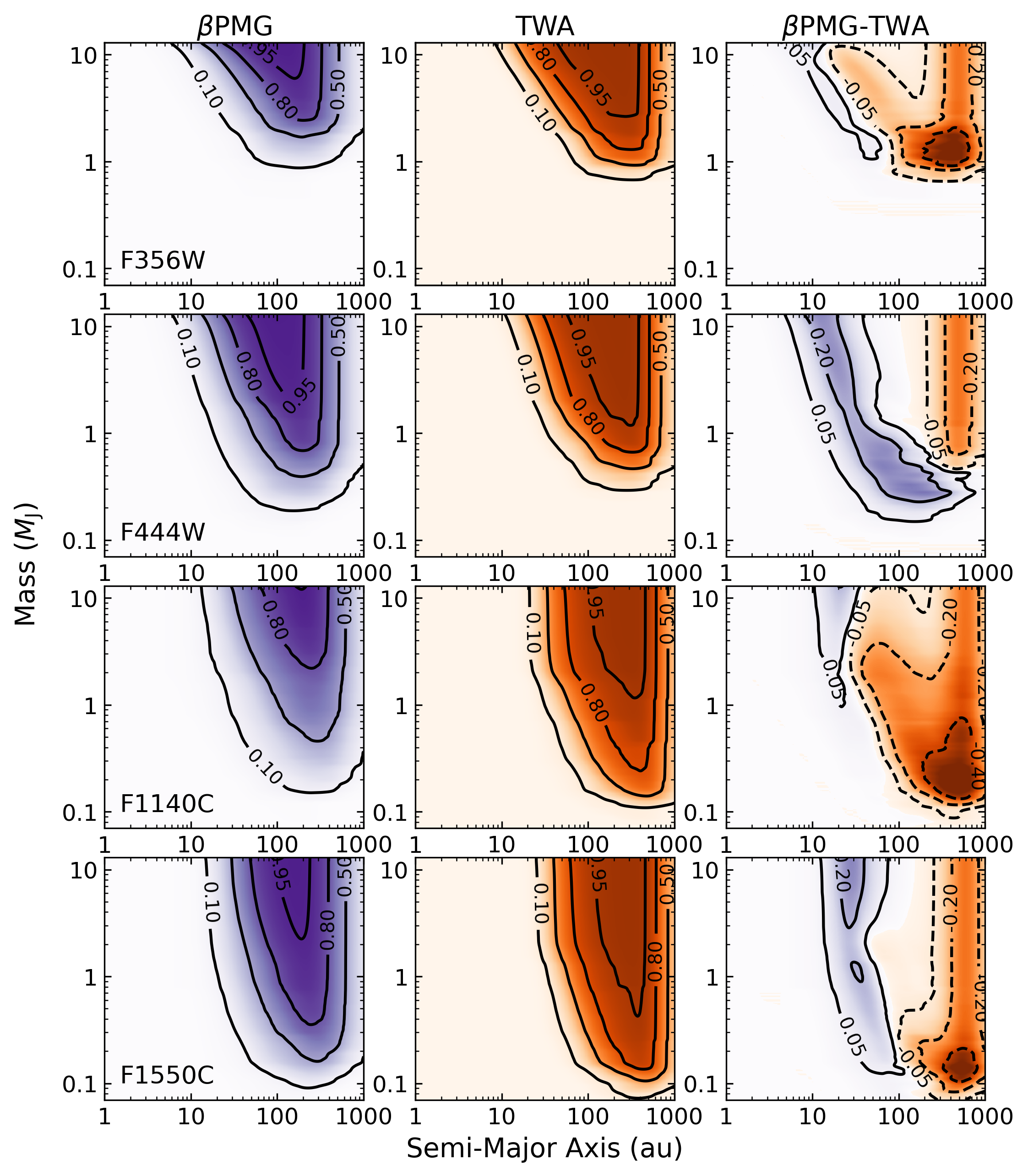}
    \caption[Moving group specific detection probability maps]{Mean detection probability maps produced using the {\sc qmess} \citep{Bona13}. From top to bottom each row corresponds to the F356W, F444W, F1140C, and F1550C \textit{JWST} filters. The first two columns correspond to the mean probability maps for all objects within the $\beta$PMG (purple) and TWA (orange) samples, contours signify the 10\%, 50\%, 80\%, and 95\% detection thresholds. The final column is equal to the difference of the $\beta$PMG and TWA columns, solid contours signify absolute detection threshold differences of 20\% and 10\%, and dashed contours signify differences of $-10$\%, $-25$\%, and $-40$\%. Irrespective of the moving group the MIRI F1140C and F1550C filters provide the best sensitivity, reaching masses as low as 0.1~$M_\textrm{J}$. Between the two moving groups, TWA is generally sensitive to the lowest mass companions at wide separations, whilst $\beta$PMG is generally more sensitive to companions at shorter separations.}
    \label{fig:maps_ymgs}
\end{figure*}

\begin{figure*}
    \centering
    \includegraphics[width=\textwidth]{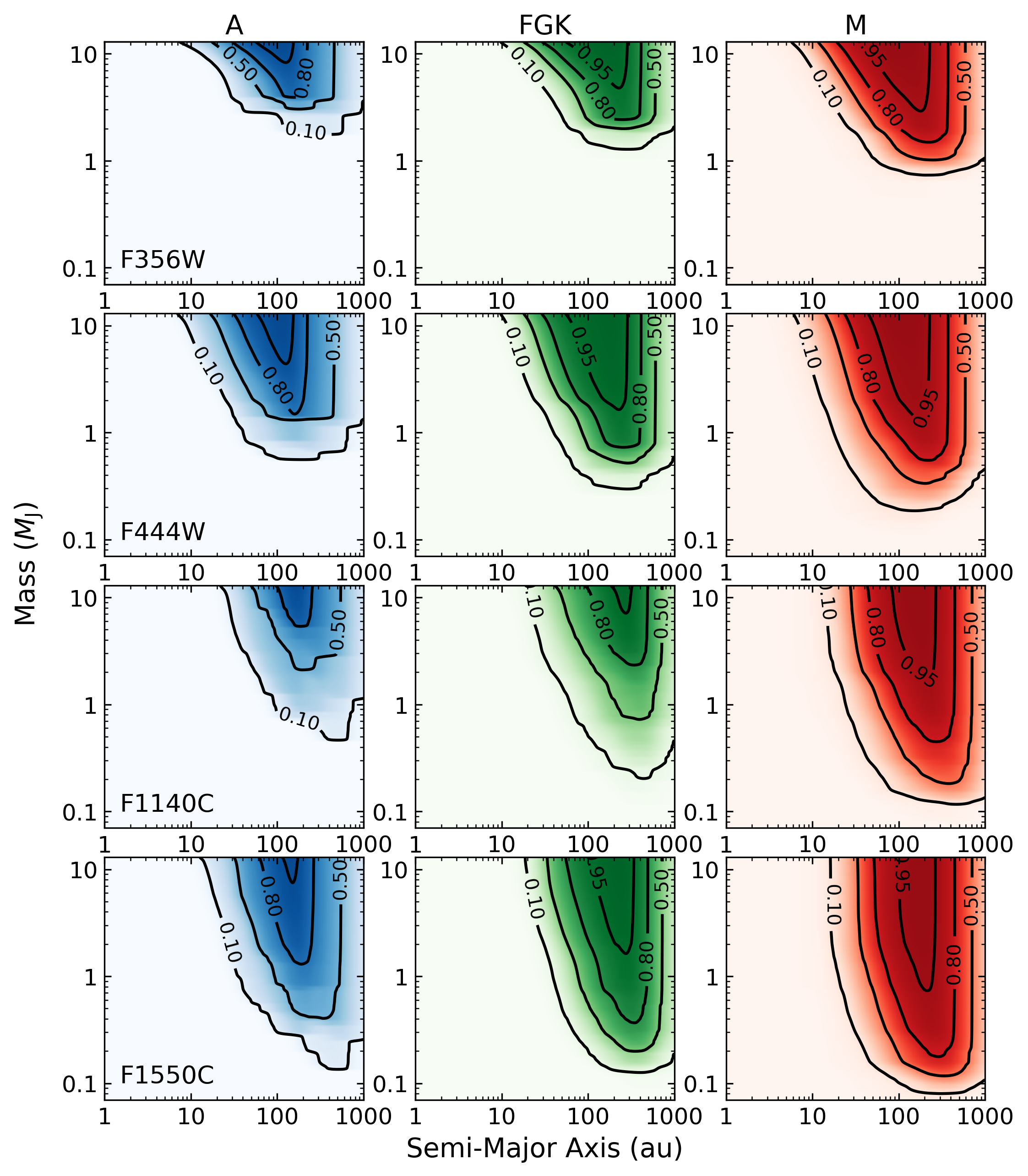}
    \caption[Spectral class specific detection probability maps]{Mean detection probability maps produced using the {\sc qmess} \citep{Bona13}. From top to bottom each row corresponds to the F356W, F444W, F1140C, and F1550C \textit{JWST} filters. From left to right each column corresponds to the mean probability maps for stars of spectral class A (blue), F/G/K (green), and M (red) across both the $\beta$PMG and TWA samples. Contours signify the 10\%, 50\%, 80\%, and 95\% detection thresholds. Similarly to \fref{fig:maps_ymgs}, the MIRI F1140C and F1550C filters provide the best sensitivity. A clear trend with sensitivity and spectral type is also observed, with M stars providing the best mass sensitivity due to their relatively fainter magnitudes than earlier spectral type objects within the moving groups.}
    \label{fig:maps_classes}
\end{figure*}

To place the broader, population based, mass sensitivity limits of \textit{JWST} in context, we present mean detection probability maps for the $\beta$PMG and TWA samples in \fref{fig:maps_ymgs}, the total combined sample separated by spectral class in \fref{fig:maps_classes}, and the total combined sample in \fref{fig:maps_full}.

\subsection{\texorpdfstring{$\beta$}{Beta} Pictoris and TW Hya}
The detection probability maps for $\beta$PMG and TWA shown in \fref{fig:maps_ymgs} clearly display an increase in sensitivity towards the lowest mass companions with increasing wavelength. In the mid-infrared the SEDs of the coolest ($\lesssim$500~K) planetary-mass objects peak, whereas a stellar SED continues to diminish. As such, the overall relative magnitude contrast between star and planet will decrease and the planet is easier to detect. This increase in detection probability for the MIRI filters therefore indicates that the favourable decrease in contrast towards mid-infrared wavelengths outweighs the difference between the NIRCam and MIRI contrast limits as shown in \fref{fig:individual_concurves_v2}. Excluding the F356W filter, all other filters are sensitive to sub-Jupiter mass exoplanets for both TWA and $\beta$BPMG. The ability to probe this parameter space is unique to \textit{JWST}, and is discussed further in Section \ref{sec:gb_comp}. The dramatic improvement between the F356W and F444W filters is primarily due to the F356W filter lying directly on a CH$_4$ absorption feature, across which significantly less flux is emitted. Comparing the measured flux between these two filters will therefore be useful in eliminating background stars, which should not exhibit such a decrease in the F356W band. In terms of the sensitivity at the shortest separations, the 50\% probability contour for the best performing F1550C filter only reaches sub-Jupiter mass companions at separations greater than $\sim$50~au for both the $\beta$PMG and TWA samples. For the same filter however, the sensitivity dramatically improves at wider separations, with the 50\% probability contour reaching a masses below $\sim$0.2~$M_\textrm{J}$ from $200-500$~au for $\beta$PMG, and from $150-800$~au for TWA. At ages of 24 and 10~Myr, corresponding to $\beta$PMG and TWA respectively, these wide separation mass limits correspond to objects with a temperature of $\sim$250~K as determined from the \texttt{BEX} evolutionary models. 

The rightmost column in \fref{fig:maps_ymgs} displays the difference in detection probabilities between the $\beta$PMG and TWA samples. Irrespective of the chosen filter, the TWA sample typically performs better at the widest separations. This is a natural result of the objects in the TWA sample being located further away on average than those in the $\beta$PMG sample (see \fref{fig:sample_hists}); the angular separations probed in an observation of a more distant object correspond to larger physical separations. Similarly, the $\beta$PMG targets perform slightly better than the TWA targets at the smallest separations as the sharp reduction in the limiting contrast at shorter angular separations occurs at a shorter physical separation for objects closer to us. In the majority of filters the TWA sample is most sensitive to the lowest mass companions due to its younger age of 10$\pm$3~Myr compared to 24$\pm$3~Myr for $\beta$PMG. At this younger age potential exoplanets will have more recently formed and will therefore be hotter and more luminous (e.g. \citealt{Bara03}, \citealt{Phil20}), making them easier to detect. Interestingly, this is not the case for the F444W filter, in which the $\beta$PMG targets perform better. This indicates that at this wavelength, for these $\beta$PMG and TWA samples, the increase in measurable flux due to a planet being physically closer to us outweighs the increase due to youth. 

\subsection{Spectral Class}
If the entire sample is instead separated in terms of spectral class, rather than young moving group membership, a similar increase in sensitivity towards the lowest masses at longer wavelengths is observed, as shown in \fref{fig:maps_classes}. For the A star sample there are slight discontinuities in the probability map resulting from the limited sample size compared to the other spectral classes (see \fref{fig:sample_hists}). Between spectral classes there is a general increase in the overall mass sensitivity at all separations for later type stars. In particular, the 50\% probability contour for the best performing F1550C filter reaches sub-Jupiter mass companions beyond $\sim$90~au for the A stars, and $\sim$40~au for the F/G/K and M stars. Furthermore, at wider separations in the same filter, the 50\% probability contour reaches a minimum mass of: $\sim$0.4~$M_\textrm{J}$ from $150-500$~au for A stars, $\sim$0.2~$M_\textrm{J}$ from $200-500$~au for F/G/K stars, and $\sim$0.15~$M_\textrm{J}$ from $100-700$~au for M stars. Given these stars are at similar distances, earlier type stars are much brighter and will impart more noise into the final image, therefore restricting the minimum detectable mass. As the vast majority of the selected TWA sample are M stars (see \fref{fig:sample_hists}), this further explains why it outperforms the $\beta$PMG sample at the widest separations in the mean detection probability maps separated by moving group (see \fref{fig:maps_ymgs}).  

Thus far, M stars have presented relatively poor targets for detecting exoplanets through direct imaging, with multiple studies indicating that at wide separations giant planets are particularly less frequent around lower mass stars \citep{Bowl16, Niel19}. However, the observations informing these studies were not sensitive to a potential population of sub-Jupiter mass companions which can be readily imaged by \textit{JWST}. Furthermore, albeit for a sample of objects at much shorter separations, the occurrence rate of sub-Jupiter mass companions as estimated from the \textit{Kepler} mission does not decline towards later spectral types \citep{Howa12, Fres13}. In these respects, M stars may represent some of the best potential targets for directly imaging the lowest mass exoplanets to date. This is distinctly different to many ground-based direct imaging observations, where M stars are typically too faint to enable precise adaptive optics wavefront corrections, resulting in sub optimal coronagraphic suppression \citep{Hard00}. In fact, \textit{JWST} Guaranteed Time Observer (GTO) program 1184 (PI: J. Schlieder) is already scheduled to perform a small NIRCam survey over nine young and nearby M stars during Cycle 1. The results from this program will not only provide valuable scientific information on the occurrence rates of sub-Jupiter mass companions at wide separations, but will also serve as a valuable comparison to the simulated sensitivities shown in this study. 

\subsection{Comparison to Ground-Based Instrumentation}\label{sec:gb_comp}
\begin{figure*}
    \centering
    \includegraphics[width=\textwidth]{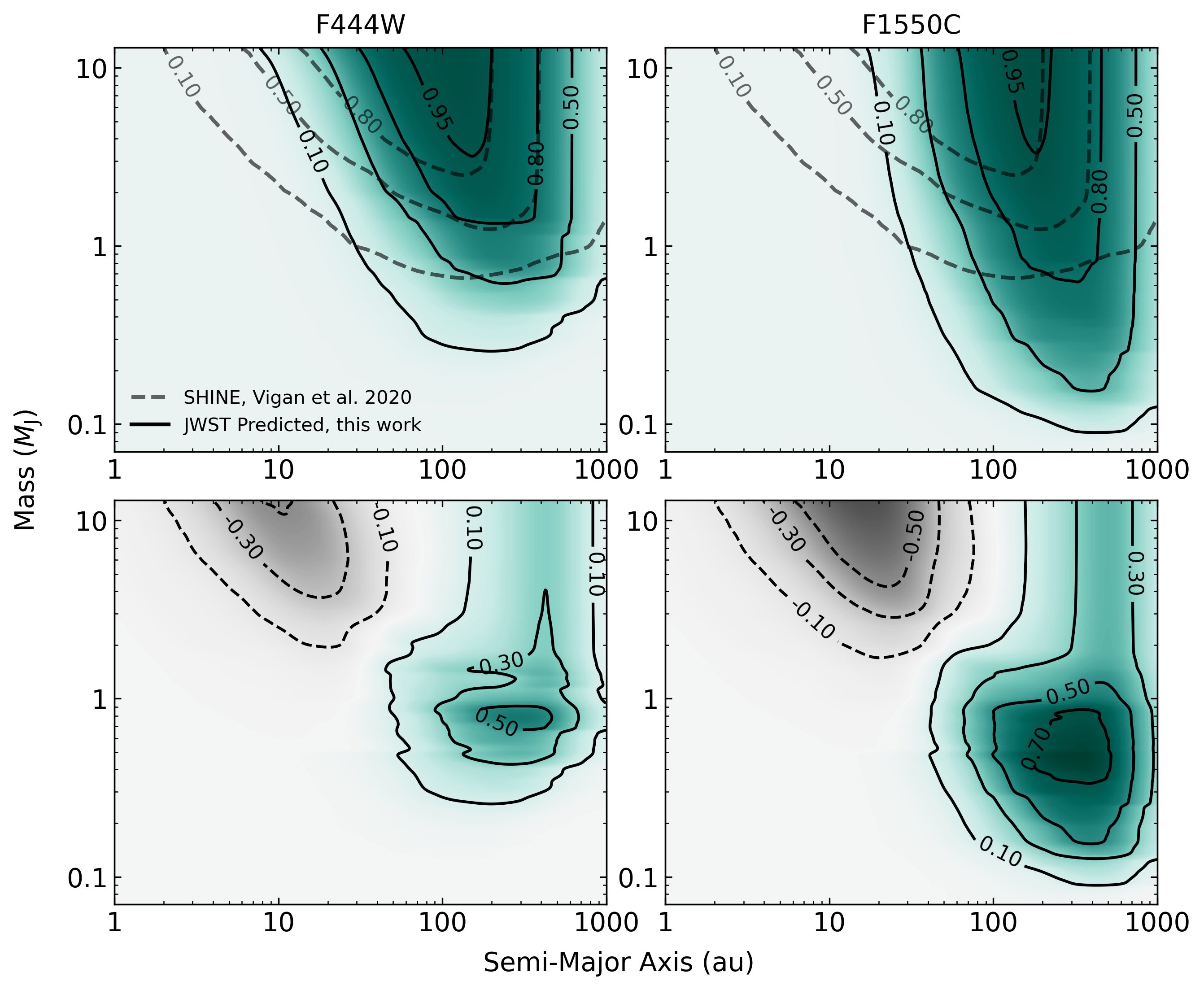}
    \caption[Full sample detection probability map comparisons]{\textit{Top}: Mean detection probability map for the full $\beta$PMG and TWA sample in the F444W (\textit{left}) and F1550C (\textit{right}) filters. Solid black contours signify the the 10\%, 50\%, 80\%, and 95\% detection thresholds of this study, and dashed grey contours signify the 10\%, 50\%, and 80\% thresholds generated from an equivalently sized sub-sample of the SHINE survey \citep{Viga20}. \textit{Bottom}: The difference between the mean detection probability map shown in the top row and the underlying mean detection probability map of the SHINE survey sub-sample contours shown in the top row. Solid contours signify absolute detection threshold differences of 70\%, 30\%, 50\% and 10\% (where \textit{JWST} is superior, teal region), and dashed contours signify absolute differences of $-10$\%, $-30$\% and $-50$\% (where SHINE is superior, grey region). Whilst \textit{JWST} offers modest improvements to direct imaging of companions at the shortest separations, beyond 30~au the sensitivity rapidly improves and sub-Jupiter mass objects are more readily detectable across a broad sample of targets.}
    \label{fig:maps_full}
\end{figure*}

To compare the predicted sensitivities of \textit{JWST} with the capabilities of a state-of-the-art ground-based instrument, the overall mean detection probability map of our entire sample in the F444W and F1550C filters, alongside mean detection probability contours generated from an equivalently sized sub-sample of the \textit{VLT} SpHere INfrared survey for Exoplanets (SHINE) \citep{Beuz19,Viga20} are shown in \fref{fig:maps_full}. In the lower panels of \fref{fig:maps_full} we additionally show the difference between this overall mean detection probability map and the underlying mean detection probability map from which the SHINE sub-sample contours were generated. The SHINE sub-sample is produced by selecting the 94 targets from the SHINE survey which have the highest total detection probabilities to an object with a mass of 2~$M_\textrm{J}$ across all separations~$-$~ensuring that only the most optimal targets are included in the overall detection probability contour determination. It is critical to note that whilst the SHINE sub-sample is identical in size to our sample, the constituent targets within each sample are different and the comparisons between their mean detection probabilities are therefore not truly one-to-one. Nevertheless, as the SHINE survey is designed to be one of the most sensitive to date, the broad improvements in mass sensitivity provided by \textit{JWST} (as described below) are predominantly due to its greater sensitivity and unique wavelength coverage, and not a result of our sample selection.

Whilst \textit{JWST} will provide very little to no improvement towards imaging the closest separation exoplanets at $\lesssim20$~au, it offers a dramatic increase in sensitivity at wider separations. Between 20 and 100~au, for the best performing F1550C filter, the detection probability increases rapidly as a function of separation, with the 50\% probability contour starting at $\sim$2~$M_\textrm{J}$ at 40~au and reaching $\sim$0.3~$M_\textrm{J}$ at 150~au. At even wider separations the detection probability improves further, reaching masses of $\sim$0.15~$M_\textrm{J}$. In contrast, the 50\% probability contour for the SHINE sub-sample starts similarly at $\sim$2~$M_\textrm{J}$ at 40~au, but only reaches $\sim$1.5~$M_\textrm{J}$ at 150~au. At separations beyond 150~au, the SHINE 50\% contour rapidly diminishes towards $>$1~$M_\textrm{J}$ at 400~au. The overall improvement of this \textit{JWST} sample compared to the SHINE sub-sample can also be visualised in terms of the absolute difference of their detection probability maps as shown in the lower panels of \fref{fig:maps_full}. In the F1550C filter, for masses <2~$M_\textrm{J}$, the SHINE sample is the most sensitive to the closest separation companions within $10$~au, from 80$-$150~au the sensitivity of each sample is essentially equivalent, and beyond 150~au the \textit{JWST} sample has an absolute increased detection probability up to $\sim$40\% as a result of the larger field of view of MIRI. Below $\sim$2~$M_\textrm{J}$ the \textit{JWST} and SHINE samples are both unable to explore separations shorter than 40~au, however at larger separations the \textit{JWST} sample is clearly superior, with an absolute increase in detection probability of at least $\sim$20\% beyond 60~au. Whilst the improvement the \textit{JWST} sample provides towards detecting sub-Jupiter mass diminishes towards both separations $\lesssim$40~au and masses $\lesssim$0.1~$M_\textrm{J}$, sub-Jupiter mass objects are largely undetectable by the SHINE sub-sample. Instead, this sensitivity to wide separation companions in the sub-Jupiter mass regime is indicative of the superior sensitivity limits of the \textit{JWST} sample itself, and demonstrates that \textit{JWST} will be uniquely capable of exploring this parameter space. This capability is further echoed in the comparable performance of other state-of-the-art instrumentation such as \textit{Spitzer} \citep{Jans15, Durk16}, \textit{Keck} NIRC2 \citep{Bowl15, Mawe19}, \textit{Subaru} SCExAO \citep{Curr19}, \textit{VLT} NaCo \citep{Viga17}, and GPI \citep{Niel19}, which although not examined in detail, are similarly limited to $\sim$1~$M_\textrm{J}$ companions. Separately, the recently commissioned \textit{VLT} NEAR instrument has demonstrated similar capabilities to that predicted for \textit{JWST}, reaching Neptune mass sensitivity at $\sim$1$\arcsec$ around $\alpha$~Cen \citep{Kasp19}. However, as 100 hours of observing time were required to reach this sensitivity for a single object, its potential for survey observations will be greatly limited.

Recent observational surveys have demonstrated that, in general, >1~$M_\textrm{J}$ planets at wide separations beyond 10~au are rare \citep{Durk16, Brya16, Viga17, Niel19, Baro19, Viga20}. Furthermore, both core accretion \citep{Poll96} and gravitational instability \citep{Boss97} population synthesis models have shown that at separations beyond 50~au, where \textit{JWST} is most sensitive, planetary-mass companions are increasingly uncommon \citep{Forg13, Forg15, Viga17, Emse20}. In this sense, the primary advantage provided by \textit{JWST} is not in the dramatic improvements in sensitivity at separations beyond 100~au, but the more modest improvements at shorter separations. At these separations sub-Jupiter, and potentially even sub-Saturn, mass exoplanets will still be detectable, although even with an optimistic assumption for the occurrence rate of a few \%, a large number of targets will be necessary to provide statistically robust constraints on their populations. Performing such a survey with \textit{JWST} could be prohibitively expensive. Even in a favourable situation, where every target has a corresponding reference five times brighter than itself, an observing program of the full TWA and $\beta$PMG sample shown here would require over 500~hours of \textit{JWST} telescope time ($\sim$8\% of the entire Cycle 1 call), primarily due to the time intensive nature of the small-grid dithered reference observations. Nevertheless, there exist a variety of options to mitigate this cost considerably, such as: reducing the sample size, reducing the exposure time, using a sparser small-grid dither pattern, selecting targets with already evident radial velocity or astrometry signals, or scheduling observations in sequence and sharing reference stars between targets in a similar fashion to the aforementioned GTO survey of nearby M stars. Such concessions may negatively impact the achievable mass sensitivity limits, which we do not explore in this work, however the flexibility in selecting a more curated survey sample as opposed to the somewhat unfocused sample described in Section \ref{sec:sed_select} will be able to counteract these effects to some degree. In addition, and perhaps most importantly, \textit{JWST} likely presents the only opportunity to explore this parameter space until the \textit{Nancy Grace Roman Space Telescope} \citep{Sper15} or the next generation of $30-40$~m ground based telescopes (e.g. \citealt{Skid15, Tama16, Fans18}) have finished construction and commissioning.

Aside from detecting new companions, \textit{JWST} will be an excellent complementary observatory to current and future ground based instruments. From the ground, large scale surveys are more realisable, and both the F444W and F1550C mass sensitivity maps shown in \fref{fig:maps_full} demonstrate that many of the detected objects from these surveys will likely be observable with \textit{JWST} also. This is particularly noteworthy as the wavelength coverage offered by \textit{JWST} is much greater than that currently offered from the ground, enabling much more detailed atmospheric characterisations of these objects. Whilst the exact sensitivity will vary depending on the filter used, the suite of near- to mid-infrared filters shown in \fref{fig:all_pces} will enable further constraints on properties such as: the abundances of carbon- and nitrogen-bearing molecular species such as NH$_3$, CH$_4$, CO, and CO$_2$; the overall atmospheric C/O, C/H, and N/H ratios, which may provide valuable clues towards the formation and migration history of an object \citep{Ober11, Madh12, Ober16, Crid20}; the presence or enhancement of molecular species generated via disequilibrium chemistry and the influence they have on our overall understanding of the atmospheric structure and dynamics (e.g. \citealt{Barm11, Skem14, Morl18, Mile20}); and the composition of cloud opacity sources and their impact on the atmospheric emission (e.g. \citealt{Morl12, Morl14, Gao20}). Many of these listed qualities are difficult, if not impossible, to characterise with current instruments owing to their limited wavelength ranges or inferior sensitivities, and \textit{JWST} direct imaging observations will therefore be crucial to advance our understanding of the widest separation exoplanets. 

%%%%%%%%%%%%%%%%%%%%%%%%%%%%%%%%%%%%%%%%%%%%%%%%%%%%%%%%%%%%%%%%%%%%%%%%%%%%%%%%%%%%%%%%%%%%%%%%%%%%%%%%%
%%% CONCLUSION
%%%%%%%%%%%%%%%%%%%%%%%%%%%%%%%%%%%%%%%%%%%%%%%%%%%%%%%%%%%%%%%%%%%%%%%%%%%%%%%%%%%%%%%%%%%%%%%%%%%%%%%%%
\section{Conclusion}\label{sec:conclusions}
We present in this work the most sophisticated simulated mass sensitivity limits for \textit{JWST} coronagraphy to date, with a particular focus on members of the nearby young moving groups TW~Hya and $\beta$~Pictoris. Of the two samples, TW~Hya members are slightly more sensitive to lower mass companions at due to their younger age, whilst $\beta$~Pictoris members are slightly more sensitive to closer separation companions because they are less distant. When separating our sample by spectral class, we find that the typically less luminous M star population provides sensitivity to the lowest mass companions. This is a stark contrast to ground-based observations, for which M stars are often too faint to facilitate the crucial adaptive optics corrections necessary for high contrast imaging \citep{Hard00}. Irrespective of spectral class or moving group, we identify the MIRI F1550C filter as the most sensitive to the lowest mass exoplanets. Across the full simulated sample, we find that \textit{JWST} will be capable of imaging $\lesssim$1~$M_\textrm{J}$ companions beyond $30$~au, $\lesssim$0.3~$M_\textrm{J}$ companions beyond $50$~au and $\sim$0.1~$M_\textrm{J}$ companions beyond 100~au. These limits represent significant improvements over surveys using current state-of-the-art ground-based instruments which are currently sensitive to $\sim$1~$M_\textrm{J}$ companions. As a result, a survey of nearby young moving group members with \textit{JWST} would be able to provide robust constraints on the presence and frequency of sub-Jupiter mass exoplanets beyond 30~au for the first time. Such measurements will be informative to planetary formation simulations, in addition to modelling of the overall population distribution. However, depending on the number of targets in the survey sample, this could be particularly time intensive. Even without such a survey, the mass sensitivity and wavelength coverage of \textit{JWST} make it an excellent tool for characterising exoplanets discovered from the ground. 

Finally, we eagerly await the launch of \textit{JWST}, at which point it will be possible to update and refine the contrast model shown in this study by the true on-sky coronagraphic observations. Such a comparison is a specific goal of the Director's Discretionary Early Release Science Program 1386, \textit{High Contrast Imaging of Exoplanets and Exoplanetary Systems with JWST} (PI: S. Hinkley), and will dramatically improve our understanding of the significance of observational factors such as pointing offset, thermal drift, target-reference slew distance, and dither strategy.

\section*{Acknowledgements}
We thank Eric Mamajek for valuable discussions on pre-main sequence stars, and Camilla Danielski for providing the MIRI coronagraphic transmission profiles. We also thank the anonymous reviewer for their useful comments on this manuscript. A significant portion of this work was performed whilst A.L.C was funded by a UK Science and Technology Facilities Council (STFC) studentship. M.B. acknowledges funding by the UK STFC grant no. ST/M001229/1. This work benefited from the 2019 Exoplanet Summer Program in the Other Worlds Laboratory (OWL) at the University of California, Santa Cruz, a program funded by the Heising-Simons Foundation. This research has made use of the NASA Astrophysics Data System and the {\sc python} modules \texttt{NumPy}, \texttt{matplotlib}, \texttt{Astropy} and \texttt{SciPy}.

\section*{Data Availability}
The data underlying this article will be shared on reasonable request to the corresponding author.

\bibliographystyle{mnras}
\bibliography{hci_jwst}

%%%%%%%%%%%%%%%%% APPENDICES %%%%%%%%%%%%%%%%%%%%%
\appendix
\section{Target Sample}

\begin{table*}
\centering 
\begin{tabular}{l c c c c c c c} 
\hline 
\hline 
\vspace{-8pt} \\ 
Common Name & 2MASS Identifier & Distance (pc) & Spectral Type & $m_\mathrm{F356W}$ & $m_\mathrm{F444W}$ & $m_\mathrm{F1140C}$ & $m_\mathrm{F1550C}$ \\ 
\hline 
\vspace{-10pt} \\ 
HD 203 & 00065008$-$2306271 & 40.0 $\pm$ 0.1 & F2IV & 5.03 & 5.03 & 5.54 & 5.49 \\ 
RBS 38 & 00172353$-$6645124 & 36.81 $\pm$ 0.04 & M2.5V & 7.59 & 7.48 & 8.19 & 7.99 \\ 
GJ 2006 A & 00275023$-$3233060 & 34.9 $\pm$ 0.1 & M3.5Ve & 7.78 & 7.65 & 8.35 & 8.16 \\ 
Barta 161 12 & 01351393$-$0712517 & 37.3 $\pm$ 0.1 & M4.3 & 7.91 & 7.78 & 8.47 & 8.28 \\ 
TYC 1208$-$468$-$1 & 01373940+1835332 & 52.1 $\pm$ 0.3 & K3Ve & 6.51 & 6.52 & 7.27 & 7.11 \\ 
HD 14082 A & 02172527+2844423 & 39.8 $\pm$ 0.1 & F5V & 4.88 & 4.88 & 5.84 & 5.78 \\ 
J0224+2031  & 02241739+2031513 & 68.7 $\pm$ 0.8 & M6 & 11.29 & 11.14 & 11.88 & 11.70 \\ 
AG Tri A & 02272924+3058246 & 41.1 $\pm$ 0.1 & K8 & 6.99 & 7.02 & 7.67 & 7.53 \\ 
EPIC 211046195 & 03350208+2342356 & 51.2 $\pm$ 0.4 & M8.5 & 10.88 & 10.73 & 11.38 & 11.20 \\ 
51 Eri & 04373613$-$0228248 & 29.8 $\pm$ 0.1 & F0V & 4.32 & 4.32 & 4.82 & 4.78 \\ 
J0443+0002  & 04433761+0002051 & 21.1 $\pm$ 0.1 & M9$\gamma$ & 10.54 & 10.40 & 10.90 & 10.75 \\ 
Gl 182 & 04593483+0147007 & 24.4 $\pm$ 0.02 & M0Ve & 6.07 & 6.04 & 6.81 & 6.64 \\ 
CD$-$57 1054 & 05004714$-$5715255 & 26.9 $\pm$ 0.02 & M0.5e & 6.05 & 6.02 & 6.81 & 6.64 \\ 
V1841 Ori & 05004928+1527006 & 53.4 $\pm$ 0.1 & K2IV & 7.54 & 7.57 & 8.23 & 8.08 \\ 
HIP 23418 ABCD & 05015881+0958587 & 23.85 $\pm$ 0.05 & M3V & 6.05 & 5.92 & 6.76 & 6.57 \\ 
J0506+0439  & 05061292+0439272 & 27.8 $\pm$ 0.04 & M4.0 & 7.86 & 7.72 & 8.42 & 8.23 \\ 
AF Lep & 05270477$-$1154033 & 26.87 $\pm$ 0.02 & F7 & 4.78 & 4.78 & 5.35 & 5.28 \\ 
J0529$-$3239  & 05294468$-$3239141 & 29.87 $\pm$ 0.04 & M4.5 & 8.05 & 7.91 & 8.60 & 8.41 \\ 
J0531$-$0303  & 05315786$-$0303367 & 38.6 $\pm$ 0.2 & M5 & 8.27 & 8.13 & 8.82 & 8.63 \\ 
V1311 Ori AB & 05320450$-$0305291 & 34.6 $\pm$ 0.7 & M2Ve & 6.82 & 6.75 & 7.47 & 7.28 \\ 
J0532$-$0301  & 05320596$-$0301159 & 38.4 $\pm$ 0.1 & M5 & 9.43 & 9.29 & 10.13 & 9.94 \\ 
Beta Pic & 05471708$-$5103594 & 19.8 $\pm$ 0.1 & A6V & 3.19 & 3.19 & 2.83 & 2.79 \\ 
GSC 06513$-$00291 & 06131330$-$2742054 & 32.7 $\pm$ 0.2 & M3.5V & 6.93 & 6.80 & 7.56 & 7.37 \\ 
AO Men & 06182824$-$7202416 & 39.26 $\pm$ 0.05 & K4Ve & 6.63 & 6.66 & 7.30 & 7.17 \\ 
TWA 22 A & 10172689$-$5354265 & 19.6 $\pm$ 0.1 & M5 & 7.39 & 7.24 & 7.96 & 7.77 \\ 
alf Cir & 14423039$-$6458305 & 15.9 $\pm$ 0.1 & A7V & 2.08 & 1.93 & 3.48 & 3.29 \\ 
V343 Nor A & 15385757$-$5742273 & 40.1 $\pm$ 0.1 & K0V & 5.42 & 5.44 & 6.33 & 6.23 \\ 
J1657$-$5343  & 16572029$-$5343316 & 50.6 $\pm$ 0.3 & M3V & 7.64 & 7.55 & 8.26 & 8.07 \\ 
HD 155555 A & 17172550$-$6657039 & 30.51 $\pm$ 0.03 & G5IV & 4.40 & 4.42 & 5.17 & 5.07 \\ 
CD$-$54 7336 & 17295506$-$5415487 & 67.8 $\pm$ 0.2 & K1V & 7.33 & 7.35 & 7.88 & 7.78 \\ 
HD 160305 & 17414903$-$5043279 & 65.7 $\pm$ 0.2 & F9V & 6.96 & 6.97 & 7.39 & 7.32 \\ 
HD 161247 & 17453733$-$2824269 & 76.2 $\pm$ 0.7 & F3V & 6.76 & 6.76 & 7.22 & 7.16 \\ 
UCAC3 74$-$428746 & 17483374$-$5306118 & 77.1 $\pm$ 0.3 & M2 & 9.11 & 9.00 & 9.71 & 9.51 \\ 
UCAC4 331$-$124196 & 17520173$-$2357571 & 63.5 $\pm$ 0.2 & M2 & 8.32 & 8.25 & 8.61 & 8.42 \\ 
HD 164249 A & 18030341$-$5138564 & 49.6 $\pm$ 0.1 & F5V & 5.71 & 5.71 & 6.16 & 6.10 \\ 
HD 165189 & 18064990$-$4325297 & 44.6 $\pm$ 0.3 & A6V & 4.14 & 4.14 & 4.68 & 4.64 \\ 
V4046 Sgr & 18141047$-$3247344 & 72.4 $\pm$ 0.3 & K6V & 7.05 & 7.08 & 5.66 & 5.52 \\ 
HD 167847 B & 18183181$-$3503026 & 83.2 $\pm$ 0.4 & G5 & 6.99 & 7.00 & 7.69 & 7.60 \\ 
HD 168210 & 18195221$-$2916327 & 79.4 $\pm$ 0.3 & G5V & 6.96 & 6.97 & 7.52 & 7.43 \\ 
J1842$-$5554  & 18420483$-$5554126 & 51.7 $\pm$ 0.2 & M4.5 & 9.54 & 9.40 & 10.13 & 9.94 \\ 
HIP 92024 A & 18452691$-$6452165 & 28.3 $\pm$ 0.2 & A7 & 4.08 & 4.08 & 3.85 & 3.81 \\ 
HD 173167 & 18480637$-$6213470 & 50.6 $\pm$ 0.1 & F5V & 6.05 & 6.05 & 6.51 & 6.44 \\ 
CD$-$31 16041 & 18504448$-$3147472 & 49.6 $\pm$ 0.1 & K7V & 7.36 & 7.35 & 8.05 & 7.88 \\ 
HIP 92680 & 18530587$-$5010499 & 47.1 $\pm$ 0.1 & G9IV & 6.23 & 6.25 & 6.84 & 6.74 \\ 
TYC 6872$-$1011$-$1 & 18580415$-$2953045 & 74.2 $\pm$ 0.4 & M0V & 7.81 & 7.80 & 8.45 & 8.28 \\ 
J1908$-$1603  & 19082195$-$1603249 & 69.4 $\pm$ 0.7 & M5.4 & 11.10 & 10.95 & 11.64 & 11.46 \\ 
HIP 95270 & 19225894$-$5432170 & 48.2 $\pm$ 0.1 & F5.5 & 5.82 & 5.82 & 6.26 & 6.20 \\ 
J1923$-$4606  & 19233820$-$4606316 & 71.1 $\pm$ 0.2 & M0V & 8.13 & 8.14 & 8.79 & 8.63 \\ 
J1935$-$2846  & 19355595$-$2846343 & 56.5 $\pm$ 1.6 & M9 & 12.02 & 11.87 & 11.32 & 11.14 \\ 
J1956$-$3207  & 19560438$-$3207376 & 51.2 $\pm$ 0.1 & M0V & 7.73 & 7.72 & 8.40 & 8.23 \\ 
HIP 99273 & 20090521$-$2613265 & 50.1 $\pm$ 0.1 & F5V & 5.89 & 5.89 & 6.37 & 6.31 \\ 
J2033$-$2556  & 20333759$-$2556521 & 43.4 $\pm$ 0.2 & M4.5V & 8.56 & 8.41 & 9.13 & 8.94 \\ 
J2043$-$2433 AB & 20434114$-$2433534 & 42.5 $\pm$ 0.2 & M3.7 & 7.53 & 7.42 & 8.18 & 7.98 \\ 
AU Mic & 20450949$-$3120266 & 9.725 $\pm$ 0.005 & M1Ve & 4.04 & 3.99 & 5.07 & 4.89 \\ 
HD 198472 & 20524162$-$5316243 & 63.1 $\pm$ 0.2 & F5.5V & 6.51 & 6.51 & 6.92 & 6.86 \\ 
HIP 103311 AB & 20554767$-$1706509 & 46.0 $\pm$ 0.1 & F8V & 5.66 & 5.67 & 6.21 & 6.14 \\ 
J2110$-$1920  & 21100461$-$1920302 & 34.3 $\pm$ 0.5 & M5 & 7.28 & 7.14 & 7.87 & 7.68 \\ 
J2135$-$4218  & 21354554$-$4218343 & 48.9 $\pm$ 0.3 & M4.5 & 10.49 & 10.34 & 11.01 & 10.82 \\ 
J2208+1144  & 22085034+1144131 & 37.0 $\pm$ 0.2 & M4.3 & 8.79 & 8.65 & 9.32 & 9.13 \\ 
HD 213429 & 22311828$-$0633183 & 25.5 $\pm$ 0.4 & F8V & 4.54 & 4.54 & 5.17 & 5.10 \\ 
CPD$-$72 2713 & 22424896$-$7142211 & 36.66 $\pm$ 0.03 & K7V & 6.73 & 6.74 & 7.44 & 7.28 \\ 
HIP 112312 A & 22445794$-$3315015 & 20.86 $\pm$ 0.02 & M4IVe & 6.66 & 6.53 & 7.31 & 7.12 \\ 
BD$-$13 6424 & 23323085$-$1215513 & 27.37 $\pm$ 0.04 & M0V & 6.40 & 6.35 & 7.15 & 6.97 \\ 
J2335$-$3401  & 23355015$-$3401477 & 38.0 $\pm$ 0.2 & M6 & 10.39 & 10.24 & 10.79 & 10.61 \\ 
\hline 
\end{tabular} 

\caption[$\beta$PMG sample properties]{Properties for all objects within the $\beta$PMG sample as obtained from the \citet{Gagn18a} compilation. The apparent magnitudes are calculated for each individual object using its corresponding synthetic SED and the \textit{JWST} PCEs as described in Section \ref{sec:simulations}.} 
\label{tab:bpmg_sample} 
\end{table*} 

\begin{table*}
\centering 
\begin{tabular}{l c c c c c c c} 
\hline 
\hline 
\vspace{-8pt} \\ 
Common Name & 2MASS Identifier & Distance (pc) & Spectral Type & $m_\mathrm{F356W}$ & $m_\mathrm{F444W}$ & $m_\mathrm{F1140C}$ & $m_\mathrm{F1550C}$ \\ 
\hline 
\vspace{-10pt} \\ 
TWA 39 A & 10120908$-$3124451 & 49.3 $\pm$ 0.4 & M4Ve & 7.62 & 7.48 & 6.93 & 6.74 \\ 
TWA 34 & 10284580$-$2830374 & 61.4 $\pm$ 0.3 & M6$\gamma$ & 9.11 & 8.96 & 9.05 & 8.86 \\ 
TWA 7 & 10423011$-$3340162 & 34.0 $\pm$ 0.1 & M4 & 6.73 & 6.64 & 7.38 & 7.19 \\ 
J1058$-$2346  & 10585054$-$2346206 & 44.1 $\pm$ 0.1 & M6$\gamma$e & 9.09 & 8.94 & 9.60 & 9.41 \\ 
TWA 1 & 11015191$-$3442170 & 60.1 $\pm$ 0.1 & M3e & 6.84 & 6.86 & 5.26 & 5.10 \\ 
TWA 43 & 11084400$-$2804504 & 53.1 $\pm$ 0.5 & A2Vn & 5.04 & 5.04 & 5.33 & 5.32 \\ 
TWA 2 A & 11091380$-$3001398 & 46.1 $\pm$ 1.4 & M1.5IVe & 6.54 & 6.47 & 7.24 & 7.05 \\ 
TWA 3 A & 11102788$-$3731520 & 36.6 $\pm$ 0.2 & M4 & 6.29 & 6.16 & 4.64 & 4.45 \\ 
TWA 12 & 11210549$-$3845163 & 65.5 $\pm$ 0.2 & M2IVe & 8.00 & 7.93 & 8.59 & 8.40 \\ 
TWA 13 A & 11211723$-$3446454 & 59.9 $\pm$ 0.1 & M1Ve & 7.52 & 7.49 & 8.14 & 7.97 \\ 
TWA 5 Aa & 11315526$-$3436272 & 49.4 $\pm$ 0.1 & M2.5 & 6.51 & 6.42 & 7.19 & 7.00 \\ 
TWA 30 A & 11321831$-$3019518 & 48.0 $\pm$ 0.3 & M5 & 8.56 & 8.42 & 7.87 & 7.68 \\ 
TWA 8 A & 11324124$-$2651559 & 46.3 $\pm$ 0.2 & M3IVe & 7.30 & 7.19 & 7.94 & 7.74 \\ 
TWA 33 & 11393382$-$3040002 & 48.7 $\pm$ 0.2 & M4.5e & 8.58 & 8.44 & 7.94 & 7.75 \\ 
TWA 26 & 11395113$-$3159214 & 49.7 $\pm$ 0.6 & M9$\gamma$ & 10.86 & 10.73 & 11.40 & 11.25 \\ 
TWA 9 A & 11482422$-$3728491 & 76.4 $\pm$ 0.4 & K7IVe & 7.64 & 7.67 & 8.28 & 8.13 \\ 
TWA 45 & 11592786$-$4510192 & 71.0 $\pm$ 1.6 & M4.5 & 8.84 & 8.70 & 9.37 & 9.18 \\ 
TWA 35 & 12002750$-$3405371 & 72.8 $\pm$ 0.5 & M4 & 8.45 & 8.31 & 9.01 & 8.82 \\ 
TWA 36 & 12023799$-$3328402 & 63.4 $\pm$ 0.3 & M5 & 9.57 & 9.43 & 10.12 & 9.93 \\ 
TWA 23 A & 12072738$-$3247002 & 55.7 $\pm$ 0.3 & M3Ve & 7.58 & 7.47 & 8.19 & 7.99 \\ 
TWA 27 A & 12073346$-$3932539 & 64.4 $\pm$ 0.7 & M8$\gamma$ & 11.13 & 10.98 & 10.29 & 10.11 \\ 
TWA 25 & 12153072$-$3948426 & 53.1 $\pm$ 0.2 & K9IV$-$Ve & 7.22 & 7.19 & 7.87 & 7.70 \\ 
TWA 44 & 12175920$-$3734433 & 76.5 $\pm$ 0.5 & M5$\gamma$e & 10.42 & 10.27 & 10.93 & 10.74 \\ 
TWA 32 A & 12265135$-$3316124 & 63.8 $\pm$ 1.4 & M5.5$\gamma$ & 9.33 & 9.18 & 8.59 & 8.40 \\ 
TWA 20 A & 12313807$-$4558593 & 81.7 $\pm$ 0.3 & M3IVe & 8.29 & 8.18 & 8.86 & 8.66 \\ 
TWA 10 & 12350424$-$4136385 & 57.6 $\pm$ 0.2 & M2Ve & 8.06 & 7.95 & 8.64 & 8.44 \\ 
TWA 46 & 12354615$-$4115531 & 56.9 $\pm$ 0.5 & M3 & 8.98 & 8.87 & 9.53 & 9.33 \\ 
TWA 11 A & 12360103$-$3952102 & 71.9 $\pm$ 0.7 & A0 & 5.40 & 5.40 & 4.97 & 4.98 \\ 
TWA 47 & 12371238$-$4021480 & 63.7 $\pm$ 0.4 & M2.5Ve & 8.36 & 8.25 & 8.94 & 8.74 \\ 
TWA 29 & 12451416$-$4429077 & 83.3 $\pm$ 3.5 & M9.5$\gamma$ & 12.77 & 12.62 & 13.88 & 13.70 \\ 
\hline 
\end{tabular} 

\caption[TWA sample properties]{As in \tref{tab:bpmg_sample}, but for the TWA sample.}
\label{tab:twa_sample} 
\end{table*} 

\section{Correlation Matrix Contrast Estimation}\label{corrmatrix_snr}

For a typical coronagraphic, or high-contrast, imaging observation it is often useful to calculate the radial contrast profile between the final noise-subtracted image and the flux of the target star. As there is usually only a single final image, this profile is typically calculated using spatial statistical properties such as the mean or variance over a range of concentric annuli. However, as the \textit{JWST} exposure time calculator \texttt{Pandeia}'s noise calculations are based on a correlation matrix infrastructure, a different method must be used for its coronagraphic simulations.

For a default \texttt{Pandeia} coronagraphic simulation, an ensemble of reference subtracted target images will be generated from a number of random draws over telescope and instrument states. An average contrast profile for these images can be determined by first calculating
\begin{equation}
    N = aCa^T,
\end{equation}
where $N$ is the noise matrix, $C$ is the covariance matrix of the ensemble of images, and $a$ is a smoothing aperture matrix. By taking the square root of the diagonal of $N$, one can obtain $\sigma$, the 2-dimensional noise map from the ensemble of images. This noise map is then normalised by the peak flux of an off-axis simulated image of the target, $S$, also smoothed by the same aperture $a$, to produce a corresponding 2-d dimensional contrast map. Finally, the radial contrast profile, or contrast curve, is determined by taking the mean within concentric annuli of the final $\sigma/S$ contrast map. 

%%%%%%%%%%%%%%%%%%%%%%%%%%%%%%%%%%%%%%%%%%%%%%%%%%
% Don't change these lines
\bsp	% typesetting comment
\label{lastpage}
\end{document}